\documentclass[aps,prd,amssymb,twocolumn,superscriptaddress,floatfix,nofootinbib,10pt]{revtex4-1}
\usepackage{latexsym}
\usepackage[normalem]{ulem}
\usepackage[utf8]{inputenc}
\usepackage{cancel}
\usepackage{graphicx}
\usepackage{color}
\usepackage{natbib}
\usepackage{float}
\usepackage{amsmath} 
\definecolor{BurntOrange}{rgb}{0.8, 0.33, 0.0}
\usepackage{hyperref}

\newcommand{\ba}{\begin{eqnarray}}
\newcommand{\ea}{\end{eqnarray}}

\newcommand{\be}{\begin{equation}}
\newcommand{\ee}{\end{equation}}

\newcommand{\im}{\textrm{Im }}

\usepackage[x11names,dvipsnames]{xcolor}

\usepackage{slashed}
\usepackage{pifont}
\usepackage{dcolumn}
\usepackage{graphics}                                                                         
\usepackage{multirow}
\usepackage{makecell}
\usepackage{bm}

\usepackage{braket}

\newcommand{\olsi}[1]{\,\overline{\!{#1}}} 

\long\def\comment#1{}

\begin{document}
                                                                  
\title{Properties of the $D_{s0}^*(2317)^\pm$ in hot and dense nuclear matter}

\author{Tomona Kinugawa}
 \email{tomona.kinugawa@riken.jp}
\affiliation{Nishina Center for Accelerator-Based Science, RIKEN, Wako 351-0198, Japan}

\author{\`Angels Ramos}
\email{ramos@fqa.ub.edu} 
\affiliation{Departament de F\'isica Qu\`antica i Astrof\'isica and Institut de Ci\`encies del Cosmos (ICCUB), Facultat de F\'isica, Universitat de Barcelona, Mart\'i i Franqu\`es 1, 08028 Barcelona, Spain}

\author{Laura Tolos}
 \email{tolos@ice.csic.es}
\affiliation{Institute of Space Sciences (ICE, CSIC), Campus UAB, Carrer de Can Magrans, 08193 Barcelona, Spain}
\affiliation{Institut d'Estudis Espacials de Catalunya (IEEC), 08860 Castelldefels (Barcelona), Spain}

\date{\today}

\begin{abstract}
We investigate the properties of the $D_{s0}^\ast(2317)^\pm$ in hot and dense nuclear matter using a coupled-channel molecular model built on next-to-leading-order heavy meson chiral perturbation theory. In-medium modifications to the $D_{s0}^\ast(2317)^+$ stem from changes to the $DK$ channel within the coupled $DK$-$D_s \eta$ system. As nuclear density increases, the $D_{s0}^\ast(2317)^+$ quasiparticle peak shifts toward lower energies and broadens, tracking the behavior of the $D$-meson spectral function. As for temperature effects, those are milder, with the thermal smearing of the Fermi surface and the melting of $\Sigma_c N^{-1}$ excitations in the $D$-meson spectral function shifting the $D_{s0}^\ast(2317)^+$ peak back toward its free-space mass while narrowing it. Conversely, the behavior of the $D_{s0}^\ast(2317)^-$ is governed by the $\bar D \bar K$ channel and its medium behavior is driven by the $\bar K$ spectral function. With increasing temperature, the $D_{s0}^\ast(2317)^-$ also approaches its free-space mass, but its width broadens before saturating at high temperatures. Incorporating explicit medium dependencies into the interaction kernel, driven by density and/or temperature variations in the pion decay constant, further shifts the $D_{s0}^\ast(2317)^+$ mass lower and narrows its width with temperature. As for $D_{s0}^\ast(2317)^-$, its mass also drops with temperature but its width increases. These contrasting medium behaviors offer a promising pathway to constrain the internal structure of these exotic states.
\end{abstract}

\maketitle


\section{Introduction}
\label{sec:intro}

The $D_{s0}^\ast(2317)^{\pm}$ is one of the first discovered exotic mesonic states, initially reported in 2003 by the BaBar Collaboration~\cite{BaBar:2003oey}, and soon after confirmed by CLEO~\cite{CLEO:2003ggt} and Belle Collaborations~\cite{Belle:2003kup}. The $D_{s0}^\ast(2317)^{\pm}$ lies far below the prediction from conventional models of $c\olsi s$ mesons~\cite{Godfrey:1985xj, Godfrey:1986wj,Zeng:1994vj,Gupta:1994mw,Ebert:1997nk,Lahde:1999ih, DiPierro:2001dwf}, while being located near the $D K$ threshold.  This state is assumed to have quantum numbers $I(J^P)=0(0^+)$ and mainly decays to the isospin-violating $D_s \pi $ channel, thus leading to a very small width ($\lesssim 3.8\,\text{MeV}$) at 95\% confidence level \cite{ParticleDataGroup:2024cfk}. 

Over the past two decades several works have discussed different possible scenarios for the nature of the $D_{s0}^*(2317)^{\pm}$. These range from conventional $c  \olsi q$ models \cite{Colangelo:2003vg, Dai:2003yg, Narison:2003td, Bardeen:2003kt, Lee:2004gt, Wang:2006bs, Lakhina:2006fy}, tetraquark $cq\olsi q \olsi q$ interactions \cite{Cheng:2003kg, Terasaki:2003qa, Chen:2004dy, Maiani:2004vq, Bracco:2005kt, Wang:2006uba}, molecular heavy-light meson-meson approaches \cite{Barnes:2003dj, Szczepaniak:2003vy, Kolomeitsev:2003ac, Hofmann:2003je, Guo:2006fu, Gamermann:2006nm, Faessler:2007gv, Flynn:2007ki, Guo:2008gp, Guo:2009ct, Liu:2012zya, Guo:2015dha, Albaladejo:2016lbb, Albaladejo:2016hae, Guo:2017jvc} to combinations of conventional quark models plus pure tetraquark or meson-meson molecules~\cite{Browder:2003fk, vanBeveren:2003kd, Ortega:2016mms, Albaladejo:2018mhb}. Also, lattice QCD has become a powerful tool to decipher the nature of the $D_{s0}^*(2317)^{\pm}$ \cite{Bali:2003jv,Dougall:2003hv,Mohler:2012na,Mohler:2013rwa,Lang:2014yfa,Bali:2017pdv,Cheung:2020mql,Alexandrou:2019tmk,Yang:2021tvc}. And very recently other methods to explore its structure have been put forward, such as the determination of the femtoscopic correlation function of the $D K$ channel \cite{Liu:2023uly,Albaladejo:2023pzq,Ikeno:2023ojl,Torres-Rincon:2023qll}.

In fact, the structure of exotic states can also be extracted by studying their behavior under extreme conditions of density and/or temperature, such as those found in heavy-ion collisions (HICs) at the Relativistic Heavy-Ion Collider (RHIC), the Large Hadron Collider (LHC), or the future Compressed Baryonic Matter (CBM) experiment at the Facility for Antiproton and Ion Research (FAIR). It is expected that compact configurations behave differently in a hot/dense medium as compared to loosely bound molecular states, as discussed  for the case of the exotic  $X(3872)$ or $T_{cc}$ states (see, for example, the discussion in Refs.~\cite{Albaladejo:2021cxj,Montesinos:2023qbx}).  

A preliminary study of the properties of $D_{s0}^\ast(2317)^+$ in dense matter was performed in Ref.~\cite{Molina:2009zeg}, based on an SU(4) extension of the SU(3) meson-meson Lagrangian. In Refs.~\cite{Montana:2020lfi,Montana:2020vjg,Montana:2023sft} the impact of a thermal medium on the $D_{s0}^\ast(2317)^+$ was analyzed. Starting from the $D K$ interaction (and coupled channels) within the next-to-leading order (NLO) of the heavy meson chiral perturbation theory (HMChPT), special attention was paid to the evolution of its mass and decay width as functions of temperature, while providing an analysis of the chiral-symmetry restoration in the heavy-flavor sector below the transition temperature. More recently,  the modifications that a dense nuclear medium induces in $D_{s0}^\ast(2317)^+$ and its antiparticle $D_{s0}^\ast(2317)^-$ have been studied in Ref.~\cite{Montesinos:2024uhq}. These states are obtained as s-wave $D K$ and $\bar D \bar K$ molecules, respectively, which are dynamically generated from effective interactions that lead to different Weinberg compositeness scenarios. Very different density patterns for $D_{s0}^\ast(2317)^+$ and $D_{s0}^\ast(2317)^-$ have been determined due to the very distinctive kaon and antikaon interactions with the nuclear medium. 

In the present paper we aim at continuing the discussion on the nature of the $D_{s0}^\ast(2317)^\pm$ by embedding them in a dense but also hot medium, thus exploring the combined effect of both density and temperature, as expected to be the case in HICs at CBM/FAIR. We start from a molecular model that couples $DK$ (${\bar D}{\bar K}$) and $D_s \eta$ (${\bar D}_s \eta$)  channels based on NLO HMChPT, as done in Refs.~\cite{Montana:2020lfi,Montana:2020vjg,Montana:2023sft}, and we investigate the changes in the $D_{s0}^\ast(2317)^\pm$ lineshapes as we consider different densities and temperatures. For the $D_{s0}^\ast(2317)^+$, the medium effects are incorporated in the $D K$ channel by modifying the $D$ meson properties, according to the results in hot dense matter of Ref.~\cite{Tolos:2007vh}. A similar procedure is performed for the antiparticle $D_{s0}^\ast(2317)^-$, but changing the ${\bar K}$ meson properties in the ${\bar D}{\bar K}$ channel with the model of Ref.~\cite{Tolos:2008di}.  We also consider the possible changes that a dense and hot medium might induce in the bare interactions by  modifying the weak decay constant appearing in the NLO HMChPT kernel. In this manner we are able to discuss the possible implications of chiral symmetry restoration in the mass and width of the $D_{s0}^\ast(2317)^\pm$ for very large densities and temperatures.

The paper is organized as follows. In Sec.~\ref{sec:formalism} we present the effective meson-meson  interaction within HMChPT at NLO and show how to obtain the $D_{s0}^\ast(2317)^+$ including medium modifications by determining the corresponding $D K$ loop function in hot dense matter. A similar procedure is presented for $D_{s0}^\ast(2317)^-$ and the corresponding ${\bar D}{\bar K}$ loop function. Moreover, we show how to include density and temperature corrections in the interaction kernel through the modification of the pion decay constant in a hot dense medium.  In Sec.~\ref{sec:freespace} we present the behavior of the $D_{s0}^\ast(2317)^\pm$ in free space by displaying the $DK$ ($\bar D \bar K$) loop functions and the related scattering amplitudes. We note that these two amplitudes are the same in free space. In Sec.~\ref{sec:medium} we analyze how the $D_{s0}^\ast(2317)^\pm$ varies for different densities and temperatures as the $DK$ and $\bar D \bar K$ loop functions are modified accordingly, whereas in Sec.~\ref{sec:medium_dependent} we display the changes in the $D_{s0}^\ast(2317)^\pm$ lineshapes when the interaction kernel is changed in matter. We finally present our conclusions in Sec.~\ref{sec:conclusions}. Appendix~\ref{sec:LECs} summarizes the low-energy constants and the expressions for the isospin coefficients, whereas Appendix~\ref{sec:loop-function-free} shows the analytic expressions of the loop function in free space.

\section{Formalism}
\label{sec:formalism}

In this section we present the theoretical formalism employed in this study. In Sec.~\ref{subsec:free-space}, we introduce the effective-theory framework based on chiral and heavy-quark spin-flavor symmetries to describe the dynamical generation of the $D_{s0}^{*}(2317)^+$ in free space via $s$-wave $DK$ and $D_s \eta$ scattering. We proceed similarly for the $D_{s0}^{*}(2317)^-$ generated from ${\bar D}{\bar K}$ and $\bar{D}_s \eta$ scattering. In Sec.~\ref{subsec:medium-effects}, we extend this framework to incorporate finite-temperature and finite-density effects in the $D K$ ($\bar D \bar K$) channels and also in the interaction kernel via the modification of the pion decay constant. Throughout this work, we assume the isospin-symmetric limit.

\subsection{Free-space scattering amplitude}
\label{subsec:free-space}

To investigate the properties of $D_{s0}^{*}(2317)^\pm$, we focus on the strangeness $S = \pm 1$ and isospin $I = 0$ sector, in which the relevant coupled channels are  $DK$ ($\bar D \bar K$) and $D_{s}\eta$ ($\bar D_s \eta$). We describe the $D_{s0}^{*}(2317)^\pm$ as a bound state dynamically generated from the coupled-channel $s$-wave $DK$-$D_s\eta$ ($\bar D \bar K$-$\bar D_s\eta$) scattering, without introducing an explicit bare $D_{s0}^{*}(2317)^\pm$ field. We employ the effective Lagrangian used in Refs.~\cite{Montana:2020vjg,Montana:2020lfi}. This Lagrangian is based on chiral and heavy-quark spin-flavor symmetries, and describes the low-energy interactions of charmed mesons ($D$ and $D^{*}$ mesons) with the pseudo-Goldstone bosons ($\pi$, $K$, $\bar{K}$, and $\eta$ mesons)~\cite{Kolomeitsev:2003ac,Lutz:2007sk,Guo:2009ct,Geng:2010vw,Abreu:2011ic}. Since we focus on the scattering relevant for $D_{s0}^{*}(2317)^\pm$, we restrict the effective Lagrangian to the terms involving the $D$ meson field:
\begin{align}
\mathcal{L}_{D\Phi} &= \mathcal{L}_{D\Phi}^{\rm LO} + \mathcal{L}^{\rm NLO}_{D\Phi}
\label{eq:LO+NLO}\\
\mathcal{L}_{D\Phi}^{\rm LO} 
&= \braket{\nabla^{\mu}D\nabla_{\mu}D^{\dag}} - m_{D}^{2}\braket{DD^{\dag}} 
\label{eq:LO} \\
\mathcal{L}^{\rm NLO}_{D\Phi} &= - h_{0}\braket{DD^{\dag}}\braket{\chi_{+}} + h_{1}\braket{D\chi_{+}D^{\dag}} \nonumber \\
&\quad + h_{2}\braket{DD^{\dag}}\braket{u^{\mu}u_{\mu}} + h_{3}\braket{Du^{\mu}u_{\mu}D^{\dag}} \nonumber \\
& \quad + h_{4}\braket{\nabla_{\mu}D\nabla_{\nu}D^{\dag}}\braket{u^{\mu}u^{\nu}} \nonumber \\
& \quad + h_{5}\braket{\nabla_{\mu}D\{u^{\mu}, u^{\nu}\}\nabla_{\nu}D^{\dag}}.
\label{eq:NLO}
\end{align}
Here, $D = (D^{0}\ D^{+}\ D^{+}_{s})$ denotes the antitriplet of pseudoscalar charmed-meson fields under light-flavor $SU(3)$. The pseudo-Goldstone boson fields are collected in the matrix $\Phi$,
\begin{equation}
\Phi = \begin{pmatrix}
\frac{1}{\sqrt{2}}\pi^{0} + \frac{1}{\sqrt{6}}\eta & \pi^{+} & K^{+} \\
\pi^{-} & -\frac{1}{\sqrt{2}}\pi^{0} + \frac{1}{\sqrt{6}}\eta & K^{0} \\
K^{-} & \bar{K}^{0} & -\sqrt{\frac{2}{3}}\eta 
\end{pmatrix} \ ,    
\end{equation}
from which the chiral building block $u_{\mu}$ is defined as
\begin{equation}
u_{\mu} = i(u^{\dag}\partial_{\mu}u - u\partial_{\mu}u^{\dag}), \quad
u = \exp\left(i\frac{\Phi}{\sqrt{2}f_{\pi}}\right), 
\end{equation}
where $f_{\pi} = 92.4$~MeV is the pion decay constant in vacuum. In this framework, $\eta_{8}$ is identified with the physical $\eta$ meson by neglecting $\eta$-$\eta'$ mixing. The covariant derivative is defined as $\nabla_{\mu}D = \partial_{\mu}D - D\Gamma_{\mu}$, where $\Gamma_{\mu} = 1/2(u^{\dag}\partial_{\mu}u + u\partial_{\mu}u^{\dag})$. The effect of explicit chiral symmetry breaking is encoded in $\chi_{+} = u^{\dag}\chi u^{\dag} + u\chi^\dag u$, where $\chi = {\rm diag}(m_{\pi}^{2},m_{\pi}^{2},2m_{K}^{2} - m_{\pi}^{2})$. The symbol $\braket{\ }$ denotes the trace in flavor space, and $\{\cdot,\cdot\}$ stands for the anticommutator. 

Then, we construct the interaction kernel from the effective Lagrangian. Since the $D_{s0}^{*}(2317)^\pm$ is described as a state generated  dynamically from $D\Phi$ scattering, we focus on the $D\Phi\to D\Phi$ interaction. The leading contribution is given by the Weinberg-Tomozawa type contact interaction arising from the covariant derivative term in the LO Lagrangian in Eq.~\eqref{eq:LO}. The NLO Lagrangian in Eq.~\eqref{eq:NLO} gives contact terms proportional to the low-energy constants $h_i$:
\begin{widetext}
\begin{align}
V_{ij}(s,t,u) &= \frac{1}{f_{\pi}^{2}} \left[\frac{C_{\rm LO}^{ij}}{4}(s - u) - 4C_{0}^{ij}h_{0} + 2C_{1}^{ij}h_{1} -2C_{24}^{ij}\left\{2h_{2}(p_{2}\cdot p_{4}) + h_{4}[(p_{1}\cdot p_{2})(p_{3}\cdot p_{4}) + (p_{1}\cdot p_{4})(p_{2}\cdot p_{3})]\right\} \right. \nonumber\\
\quad & \quad \quad \quad \left. + 2C_{35}^{ij}\{h_{3}(p_{2}\cdot p_{4}) + h_{5}[(p_{1}\cdot p_{2})(p_{3}\cdot p_{4}) + (p_{1}\cdot p_{4})(p_{2}\cdot p_{3})]\}\right].
\label{eq:V}
\end{align}
\end{widetext}
Here $s = (p_{1} + p_{2})^{2}$, $t = (p_{1} - p_{3})^{2}$, and $u = (p_{1} - p_{4})^{2}$ are the Mandelstam variables. We denote the four-momenta of the incoming mesons by $p_1$ and $p_2$, and those of the outgoing ones by $p_3$ and $p_4$, respectively. For the numerical calculation, we use the low-energy constants $h_i$ determined from lattice QCD data in Ref.~\cite{Guo:2018tjx}. The numerical values of these low-energy constants and the explicit expressions for the coefficients $C_{{\rm LO},0,1,24,35}$ in the isospin basis are summarized in Appendix~\ref{sec:LECs}.

Using the relations among the Mandelstam variables, together with the on-shell conditions $p_i^2=m_i^2$, Eq.~\eqref{eq:V} can be expressed in terms of $s$ and the scattering angle. Since we focus only on $s$-wave scattering, we perform the $s$-wave projection. As a result, the interaction kernel is reduced to a function of $s$ alone, denoted by $V_{ij}(s)$. 

The interaction kernel constructed above is used to obtain the $D\Phi$ scattering amplitude. In this work, we employ the on-shell approximation of the amplitudes in the Lippmann-Schwinger (LS) equation, in which case it reduces to an algebraic one. Then the coupled-channel $T$-matrix is given by
\begin{align}
T(s) = \left[V^{-1}(s)-G(s)\right]^{-1},
\label{eq:Tmatrix}
\end{align}
where $T(s)$, $V(s)$, and $G(s)$ are $2\times2$ matrices, with $V(s)$ constructed from the kernels $V_{ij}(s)$ obtained above. Here, $G(s)$ is the diagonal matrix of the two-meson loop functions. Its diagonal component for a given channel $D\Phi$ is obtained as
\begin{align}
G_{D\Phi}(s) &= i \int \frac{d^{4}q}{(2\pi)^{4}} \frac{1}{q^{2} - m_{D}^{2} + i\varepsilon}\frac{1}{(P - q)^{2} - m_{\Phi}^{2} + i\varepsilon},
\label{eq:G-free}
\end{align}
where $q$ is the loop four-momentum and $\varepsilon$ is a positive infinitesimal quantity. The quantity $P = p_{1} + p_{2} = p_{3} + p_{4}$ is the total four-momentum in the two-meson system satisfying $P^{2} = s$. The loop function diverges if we evaluate the momentum integration directly, and therefore, a regularization procedure is required. In this work, we employ either a sharp cutoff or a dimensional regularization (DR) scheme. The analytic expressions of the regularized loop functions are given in Appendix~\ref{sec:loop-function-free}. The regularization parameters in each scheme are fixed so that the vacuum $D\Phi$ scattering amplitude has a pole at the mass of the $D_{s0}^{*}(2317)$. Their numerical values are summarized in Appendix~\ref{sec:LECs}.

\subsection{Medium effects and finite-temperature corrections}
\label{subsec:medium-effects}

The formulation described so far can be applied to the vacuum case, where both
the density $\rho$ and temperature $T$ are set to zero. In this section, we extend the framework to include medium effects in dense and hot matter. We first employ the straightforward approach of incorporating finite density and temperature effects into the two-meson loop function. As a possible source of medium dependence in the interaction kernel, we then consider a density and temperature-dependent pion decay constant $f_{\pi}$.

\subsubsection{In-medium and finite-temperature loop function}
\label{subsubsec:medium-loop}

Finite-temperature and finite-density effects can be incorporated into the two-meson loop functions through the thermal statistical factors and the in-medium modification of the meson propagators. At finite temperature, the two-meson loop function is evaluated within the imaginary-time formalism by replacing the integration over the internal energy with a sum over bosonic Matsubara frequencies~\cite{Matsubara:1955ws}. Performing the Matsubara summation provides terms containing the Bose-Einstein distribution function
\begin{align}
f(\omega, T) &= \frac{1}{e^{\omega/T} - 1},
\end{align}
which introduces the explicit temperature dependence in the loop function. Here $\omega$ is the energy of the meson and $T$ is the temperature.
In addition, mesons propagating in a medium at finite temperature and density can acquire in-medium self-energies through their interactions with the surrounding particles. The in-medium self-energy $\Pi_{M}$ is related to the spectral function $S_{M}$ as
\begin{align}
S_{M}(\omega, \bm{q}; T, \rho) &= -\frac{1}{\pi} \im\frac{1}{\omega^{2} - \bm{q}^{2} - m_{M}^{2} - \Pi_{M}(\omega, \bm{q}; T, \rho)},
\end{align}
where $M$ stands for the meson species. For a meson in vacuum, the spectral function reduces to a delta-function, whereas in the medium the quasiparticle peak can exhibit a shift and a broadening. By employing the Lehmann representation of the in-medium propagators in terms of their spectral functions~\cite{Kapusta:2006pm}, the in-medium modifications of the mesons can be incorporated into the two-meson loop functions~\cite{Montana:2020lfi,Montana:2020vjg}. 

In free space, the loop function of Eq.~\eqref{eq:G-free} can be written only in terms of $P^{2} = s$, because of Lorentz invariance. In a medium, however, the medium itself defines a preferred rest frame, and the two-meson loop function generally depends separately on the total energy $E$ and three-momentum $\bm{P}$. In this study, we evaluate the loop functions in the two-meson center-of-mass frame, where $\bm{P}=\bm{0}$. To simplify the notation, we denote the resulting loop functions by $G_{D\Phi}(E;T,\rho)$ with $E=\sqrt{s}$ in this frame.

In the case of the $D_{s0}^{*}(2317)^+$, the $DK$ channel includes the in-medium dressing only for the $D$ meson and assumes the $K$ meson to be an undressed particle. This approximation is motivated by the small modification of the $K$ spectral function in nuclear matter due to the mild repulsive $KN$ interaction~\cite{Tolos:2008di}. The $DK$ loop function is then given by~\cite{Montana:2020vjg}
\begin{eqnarray}
&&G_{DK}(E; T, \rho) = - \int \frac{d^{3}q}{(2\pi)^{3}}\int_{\omega_{\rm min}}^{\omega_{\rm max}} d\omega
\frac{S_{D}(\omega, \bm{q}; T, \rho)}{2 \omega_{K}} \nonumber \\
&& \times \biggl\{[1 + f(\omega, T) + f(\omega_{K}, T)] \nonumber \\
&& \quad \times \left(\frac{1}{\omega - (E - \omega_{K}) - i\varepsilon} + \frac{1}{\omega + (E + \omega_{K}) + i\varepsilon} \right) \nonumber \\
&& \quad + [f(\omega_{K}, T) - f(\omega, T)] \nonumber \\
&& \quad \left. \times \left(\frac{1}{\omega - (E + \omega_{K}) - i\varepsilon} + \frac{1}{\omega + (E - \omega_{K}) + i\varepsilon} \right)
\right\}, \hspace{0.7cm}
\label{eq:GDK-mid}
\end{eqnarray}
where $\omega_{M} = \sqrt{\bm{q}^{2} + m_{M}^{2}}$ with three-momentum $\bm{q}$. 
In principle, the energy integration is performed over the full range of the energy $\omega$. For the numerical calculation, however, we restrict the integration to $\omega_{\rm min} \leq \omega \leq \omega_{\rm max}$, chosen to cover the region where the spectral function has non-negligible strength. The values of $\omega_{\rm min}$ and $\omega_{\rm max}$ used in the calculations are given in Appendix~\ref{sec:LECs}. In evaluating Eq.~\eqref{eq:GDK-mid}, we use the $D$-meson spectral function in nuclear matter obtained in Ref.~\cite{Tolos:2007vh}.

 For the $D_s\eta$ channel, we neglect the in-medium dressing of both the $D_s$ and $\eta$ mesons, consistently with the assumptions made in the calculation for the $D$-meson spectral functions employed here \cite{Tolos:2007vh}.
Thus, the $D_s\eta$ loop function incorporates only the temperature dependence in the Bose-Einstein distribution functions and reads:
\begin{eqnarray}
&&G_{D_{s}\eta}(E; T) = \int \frac{d^{3}q}{(2\pi)^{3}} \frac{1}{4\omega_{D_{s}}\omega_{\eta}} \nonumber \\
&& \times \biggl\{ [1 + f(\omega_{D_{s}},T) + f(\omega_{\eta},T)] \nonumber \\
&& \quad \times \left(\frac{1}{E - \omega_{D_{s}} - \omega_{\eta} + i\varepsilon} - \frac{1}{E + \omega_{D_{s}} + \omega_{\eta} + i\varepsilon}\right) \nonumber \\
&& \quad + [f(\omega_{D_{s}},T) - f(\omega_{\eta},T)]\nonumber \\
&& \quad \left. \times \left(\frac{1}{E + \omega_{D_{s}} - \omega_{\eta} + i\varepsilon} - \frac{1}{E - \omega_{D_{s}} + \omega_{\eta} + i\varepsilon}\right) \right\} .
 \hspace{0.7cm} \label{eq:GDseta-mid}
\end{eqnarray}

In this way, the $DK$ loop function contains $T$ and $\rho$ dependences through the in-medium $D$-meson spectral function, together with the Bose-Einstein distribution functions. In contrast, the $D_s\eta$ loop function contains only the temperature dependence.

We recall that some studies of the $\eta$-meson in matter \cite{Waas:1997pe,Inoue:2002xw,Post:2003hu}
found its spectral function to show a narrow quasiparticle peak, which moves below the $\eta$-meson free mass and broadens as density increases, together with a bump at higher energy tied to the coupling to the $N^*(1535)N^{-1}$ excitation mode. A similar behavior is found for the $D_s$ spectral function in the study of Ref.~\cite{Jimenez-Tejero:2011dif}. Apart from the quasiparticle peak, there is a small structure to its right that reflects the enhanced cusp found in the $D_sN$ amplitude at the $K\Sigma_c$ threshold \cite{Jimenez-Tejero:2009cyn}. This mode is more visible in the $D_s$ spectral function obtained in Ref.~\cite{Lutz:2005vx}, since the model employed there~\cite{Hofmann:2005sw} produces a resonance, not a cusp, about 15 MeV below the $K\Sigma_c$ threshold. In addition, there appears a subthreshold mode, located around 75 MeV below the $D_s$ mass, related to an additional $D_s N$ resonance found in \cite{Hofmann:2005sw} and not in \cite{Jimenez-Tejero:2009cyn}, in spite of the similar kernels employed, which include a null or negligible $D_sN$ interaction. It may well be that this additional resonance is in fact a spurious pole tied to the DR scheme employed in \cite{Hofmann:2005sw} in contrast to the cutoff approach employed in \cite{Jimenez-Tejero:2009cyn}, a phenomenon recently discussed in \cite{Garcia-Gonzales:2026dnn}. Ignoring this mode, we can summarize the various findings in the literature by stating that the $\eta$ and $D_s$ spectral functions in nuclear matter present well-distinguished quasiparticle peaks, a fact that justifies our choice of implementing only the temperature effects for these mesons. However, the widening of the quasiparticle peak with density and the appearance of strength associated with genuine resonant-hole modes make it advisable to include these density dressing effects in a future calculation.

In the case of the  $D_{s0}^{*}(2317)^-$ state, we consider the $\bar D \bar K$ and $\bar D_s \eta$ loop functions. 
For the $\bar D_s \eta$ loop, we follow the same procedure as for the $D_s \eta$ loop and we only incorporate the temperature dependence implemented by the Bose-Einstein distributions, so we use the same expression as in Eq.~(\ref{eq:GDseta-mid}). For the $\bar D \bar K$ loop we implement the spectral function of the $\bar K$ meson in a hot and dense medium as it experiences strong modifications (see, for example, the discussion in the review paper of Ref.~\cite{Tolos:2020aln}). We will employ the $\bar K$-meson spectral function obtained in Ref.~\cite{Tolos:2008di}. As for the $\bar D$ meson, we recall that the work of Ref.~\cite{Tolos:2007vh} not only obtained the $D$-meson spectral function in a hot and dense medium but also derived the $\bar D$ properties. In agreement with what was also found in Ref.~\cite{Lutz:2005vx}, the $\bar D$ meson in nuclear matter at saturation density acquires a delta-type spectral shape with the quasiparticle peak slightly shifted to higher energies due to the mild and repulsive character of the $\bar D N$ interaction. Consistently with the treatment of the $K$ meson in the case of the $DK$ loop for the $D_{s0}^{*}(2317)^+$ state, we do not implement density modifications in the $\bar D$ meson. Therefore, the $\bar D \bar K$ loop is obtained from Eq.~(\ref{eq:GDK-mid}) but by exchanging the roles of the $D$ and $K$ mesons with those of the $\bar K$ and $\bar D$ mesons, respectively.

In the cutoff regularization scheme, the in-medium loop function is obtained by setting the upper limit of the three-momentum integral to a finite cutoff $\Lambda$. We employ the same value of $\Lambda$ as that in free space.
On the other hand, within the DR scheme, the in-medium loop function cannot be written analytically. Therefore, in order to incorporate the finite-density and finite-temperature effects into the DR loop function, we use the following
prescription:
\begin{equation}
 G^{\rm DR,med}_{D\Phi}(s; T, \rho) = G^{\rm DR,free}_{D\Phi}(s) + \Delta^{\rm cut} ,
\end{equation}
with
\begin{equation}
    \Delta^{\rm cut} \equiv G^{\rm cut,med}_{D\Phi}(s; T, \rho) - G^{\rm cut,free}_{D\Phi}(s).   
\end{equation}
Here, $G^{\rm DR,med}_{D\Phi}(s; T, \rho)$ denotes the in-medium loop function in DR defined by this prescription, $G^{\rm DR,free}_{D\Phi}(s)$ is the free-space DR loop function, and $G^{\rm cut,med}_{D\Phi}(s; T, \rho)$ and $G^{\rm cut,free}_{D\Phi}(s)$ are the cutoff-regularized loop functions in medium and in free space, respectively. This prescription is constructed so that when $\rho \to 0$ and $T \to 0$, $G^{\rm DR, med}_{D\Phi}(s; T, \rho)$ reduces to the DR loop function in free space.

\subsubsection{Density and temperature dependent pion decay constant}

So far, we have not accounted for the possible density and temperature dependence of the interaction kernel. In our approach, the interaction is proportional to the universal vector meson coupling constant $g$, related to the pion decay constant by $g=m_V/(2 f_\pi)$, where $m_V$ is an appropriate vector meson mass, such as that of the $\rho$ meson. Therefore, ignoring the possible density and temperature dependence of $m_V$, the behavior of the interaction in a hot and dense medium will be determined by that of $f_\pi$. The evolution of $f_\pi$ with density and/or temperature is connected to the behavior of the chiral condensate $\langle \bar{q}q  \rangle$ through the Gell-Mann-Oakes-Renner (GOR) relation:
\begin{equation}
f_\pi^2 m_\pi^2 = -(m_u + m_d) \langle \bar{q}q  \rangle \ ,
\end{equation}
and it has been thoroughly studied in the literature~\cite{Gasser:1986vb,Gerber:1988tt,Kodama:1995kj,Bochkarev:1995gi,Kim:2003tp,Kaiser:2007nv,Jido:2008bk,Alarcon:2011zs,Goda:2013npa}. 

We will absorb both the density and temperature effects in the following expression:
\begin{equation}
    f_\pi(T,\rho)=f_\pi \left[1-\frac{\sigma_{\pi N}}{2m_\pi^2 f_\pi^2}\rho \right]\left[1-\frac{1}{12}\frac{T^2}{f_\pi^2}\right] \ ,
    \label{eq:f_pi_T_rho}
\end{equation}
where $\sigma_{\pi N}$ is the pion-nucleon sigma term. The above expression combines the low temperature result in 
chiral perturbation theory for $N_f=2$ in Ref.~\cite{Gasser:1986vb} with the lowest term in the density expansion of the chiral condensate in the chiral limit~\cite{Goda:2013npa}, where $\sigma_{\pi N}$ ranges from 45 MeV to 60 MeV. We adopt $\sigma_{\pi N} = 50$~MeV for most calculations, 
but we will also present results employing the extreme values of the range for comparison.


\begin{figure*}
    \centering
\includegraphics[width=0.45\linewidth]
{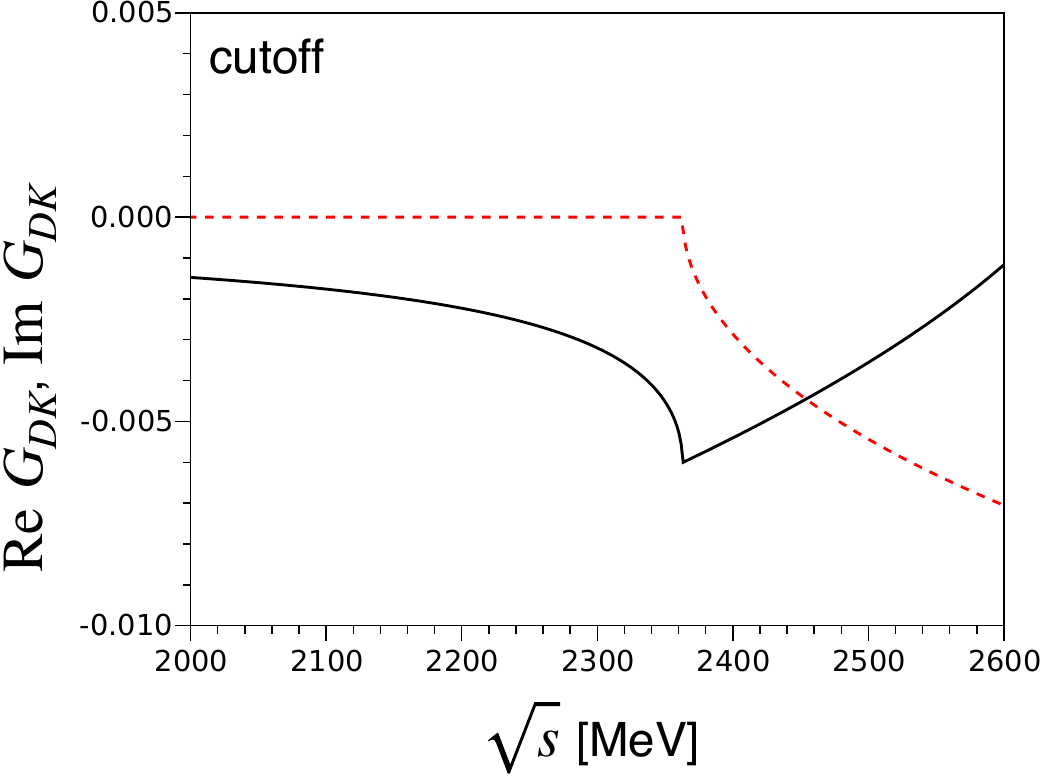}
\includegraphics[width=0.45\linewidth]
{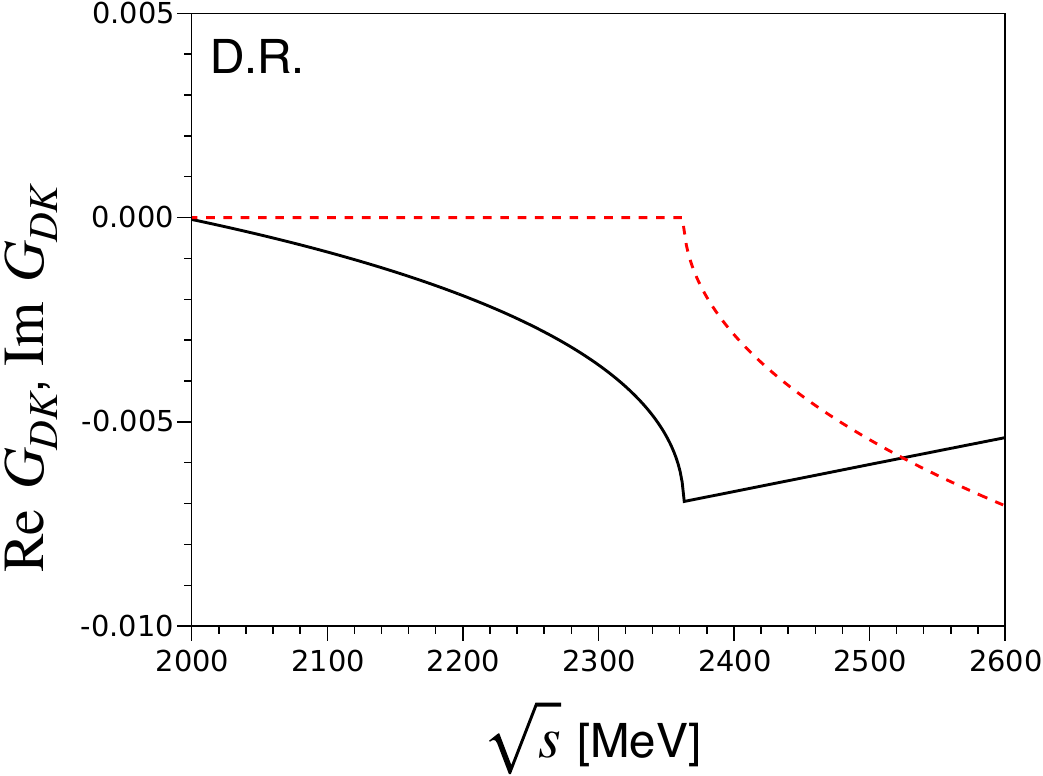}
    \caption{Real (solid lines) and imaginary (dashed lines) parts of the $DK$ loop function $G_{DK}$ in free space as functions of the center-of-mass energy $\sqrt{s}$. The left plot corresponds to the calculation of the loop function with a cutoff of $\Lambda = 615$~MeV, whereas the loop function in the right plot is computed using DR with a subtraction constant of $a = -1.79$ at a regularization scale of $\mu = 1000$~MeV.}
    \label{fig:GDK-free}
\end{figure*}

\begin{figure*}
    \centering
\includegraphics[width=0.45\linewidth]
{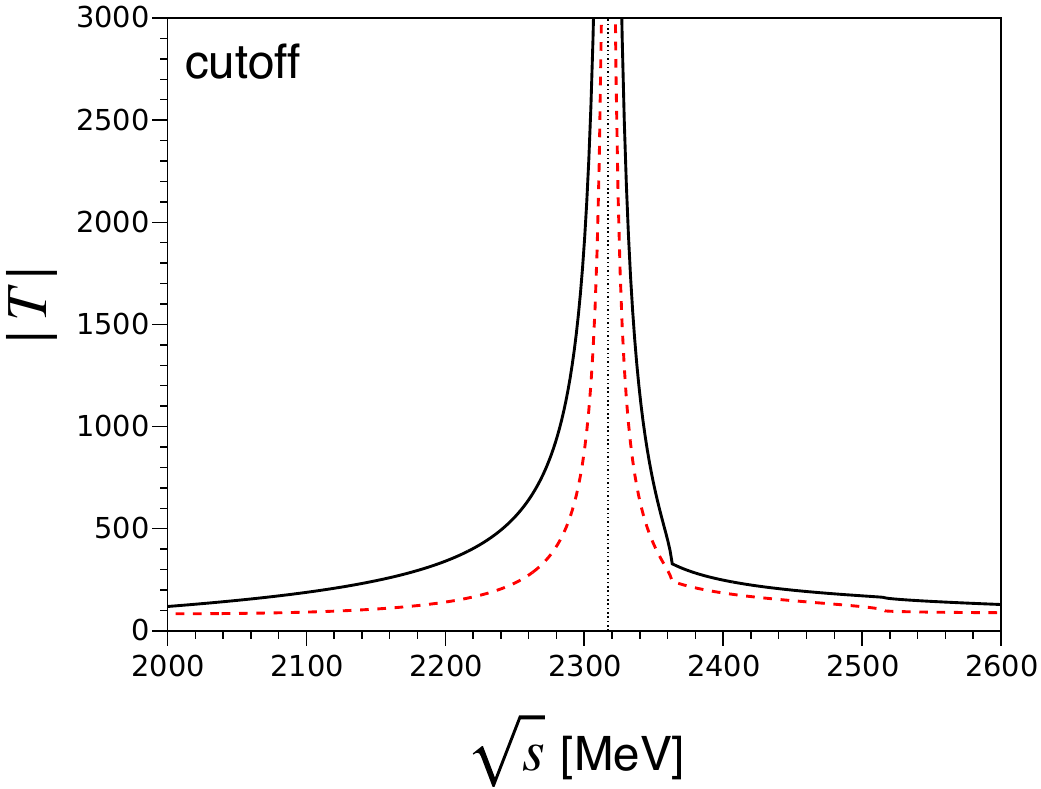}
\includegraphics[width=0.45\linewidth]
{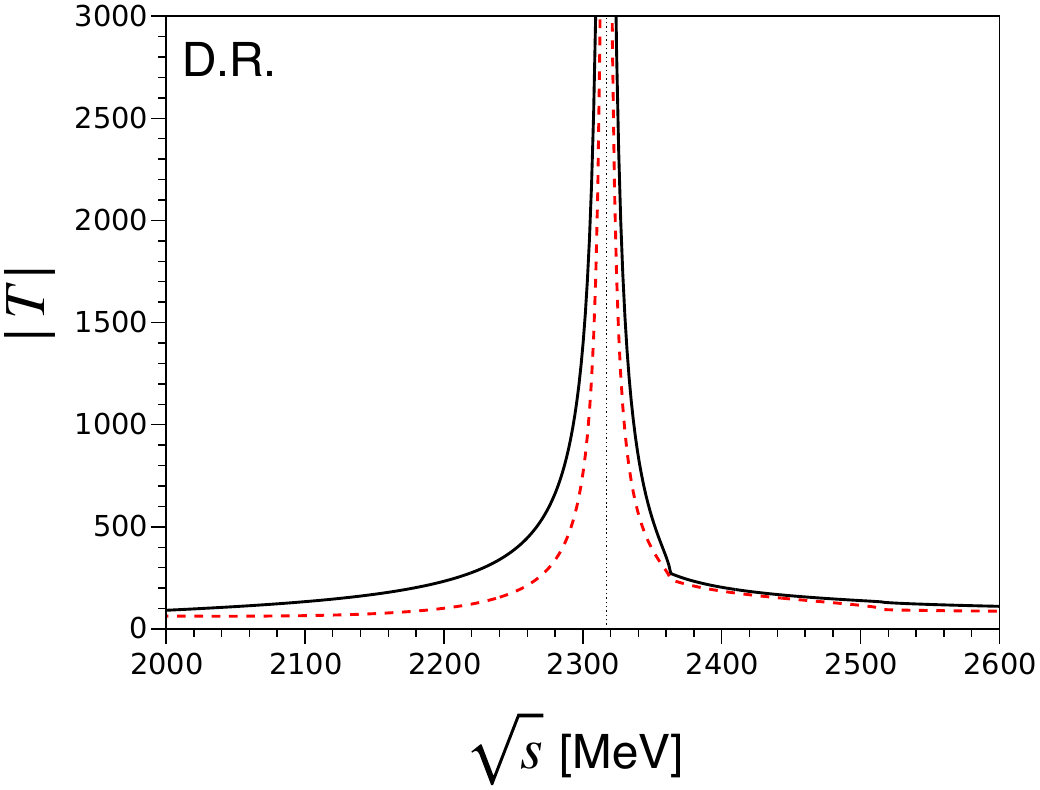}
    \caption{The modulus of the $DK$ scattering amplitude $|T_{DK}|$ in free space (black solid line) as a function of the center-of-mass energy $\sqrt{s}$ for the cutoff regularization scheme (left plot) and DR (right plot). The red dashed line in each panel displays the modulus of the $D_s\eta$ amplitude for comparison.}
    \label{fig:TDK-free}
\end{figure*}

\section{Free-Space Results}
\label{sec:freespace}

In this section we analyze the behavior of 
the two-meson loop function and scattering amplitude in free space (vacuum), where the density $\rho$ and temperature $T$ are both zero. Fig.~\ref{fig:GDK-free} displays the $DK$ loop function in free space evaluated using the sharp cutoff regularization scheme (left panel) and the DR scheme (right panel). The regularization parameters have been adjusted to reproduce the empirical mass of the $D_{s0}^*(2317)^+$ state. The resulting values are $\Lambda = 615$~MeV for the cutoff scheme, and a subtraction constant of $a = -1.79$ for the DR scheme at a regularization scale of $\mu = 1000$~MeV. Utilizing these parameters, the modulus of the $DK$ $T$-matrix as a function of the center-of-mass energy $\sqrt{s}$ is plotted in Fig.~\ref{fig:TDK-free} for both regularization prescriptions, showing a clear pole corresponding to the dynamically generated bound state.  We note that the study of the $D_{s0}^*(2317)^-$ state in free space requires the calculation of the  $\bar D \bar K$ loop function and the corresponding scattering amplitude, which are the same as those for $D K$ because of charge conjugation.

\begin{figure*}
    \centering
\includegraphics[width=0.45\linewidth]
{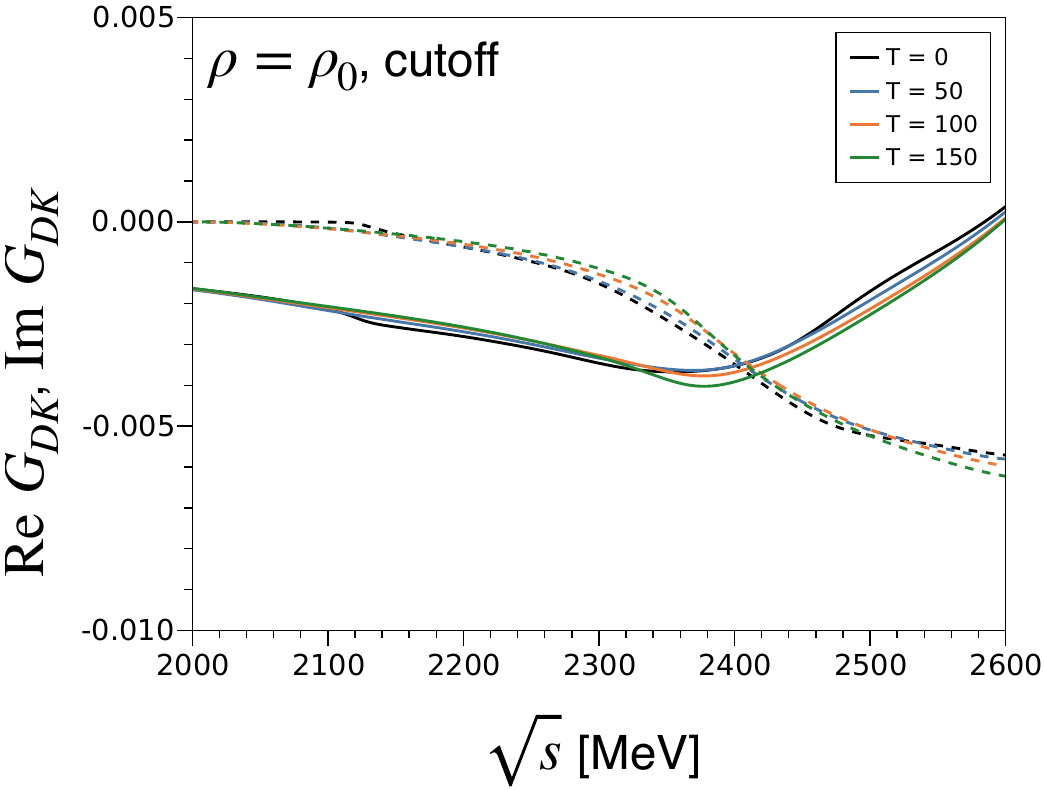}
\includegraphics[width=0.45\linewidth]
{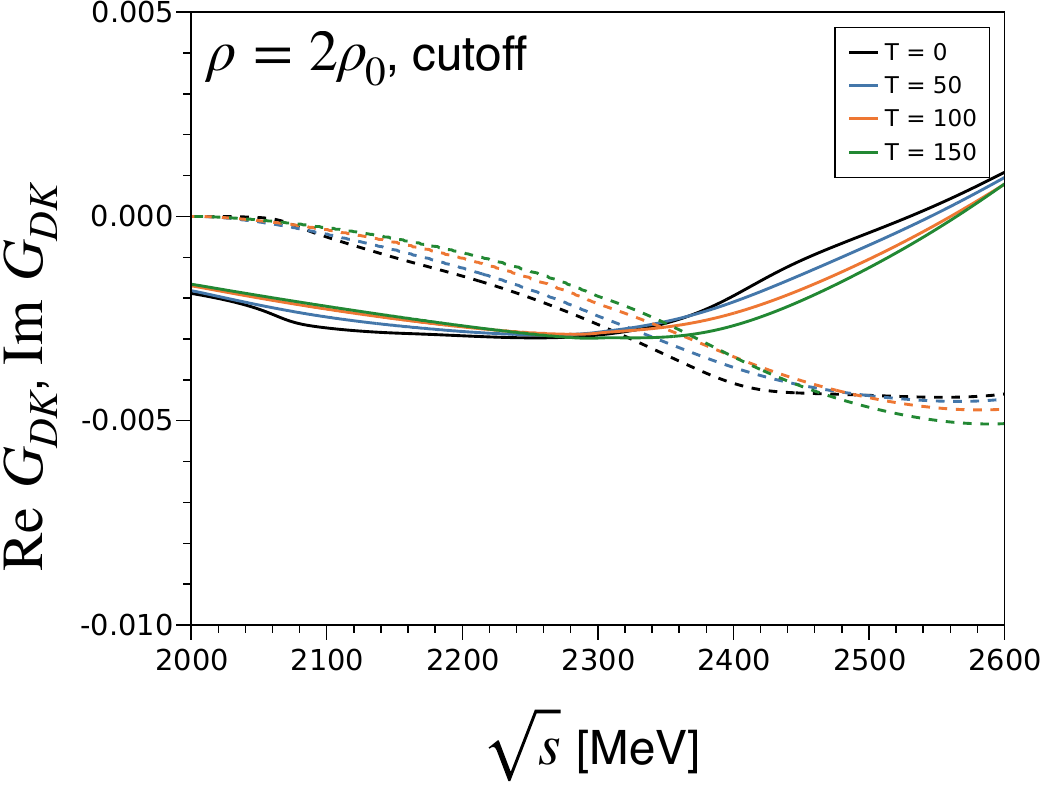}
\includegraphics[width=0.45\linewidth]
{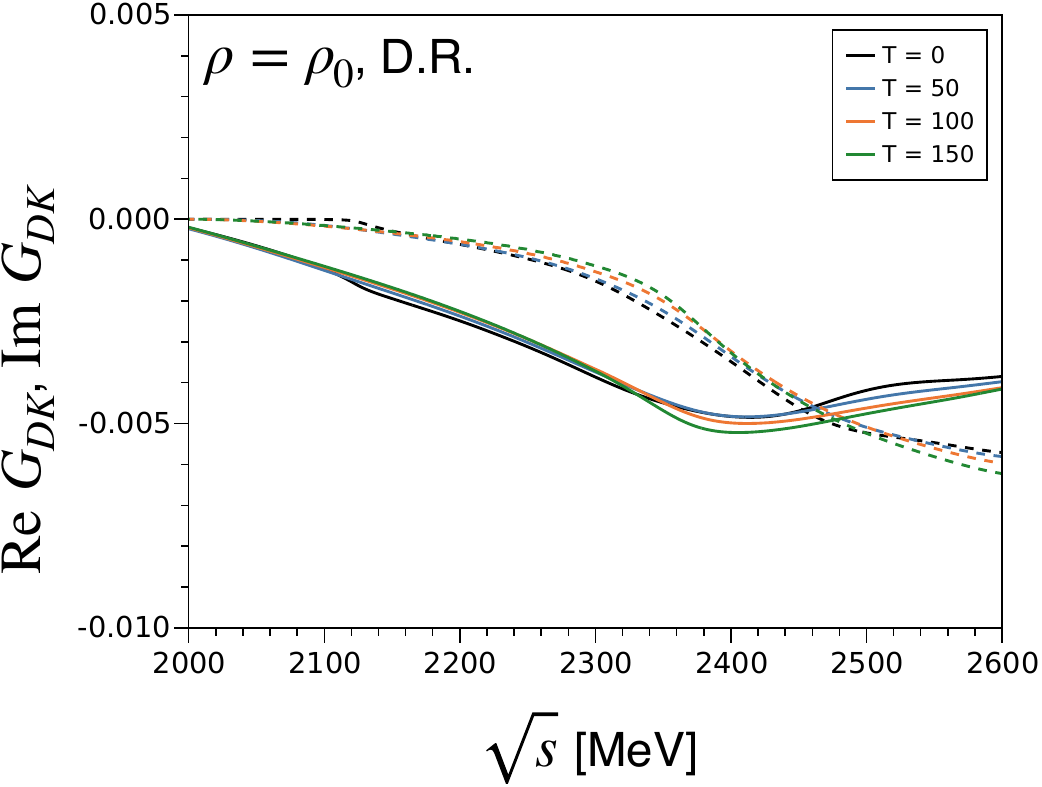}
\includegraphics[width=0.45\linewidth]
{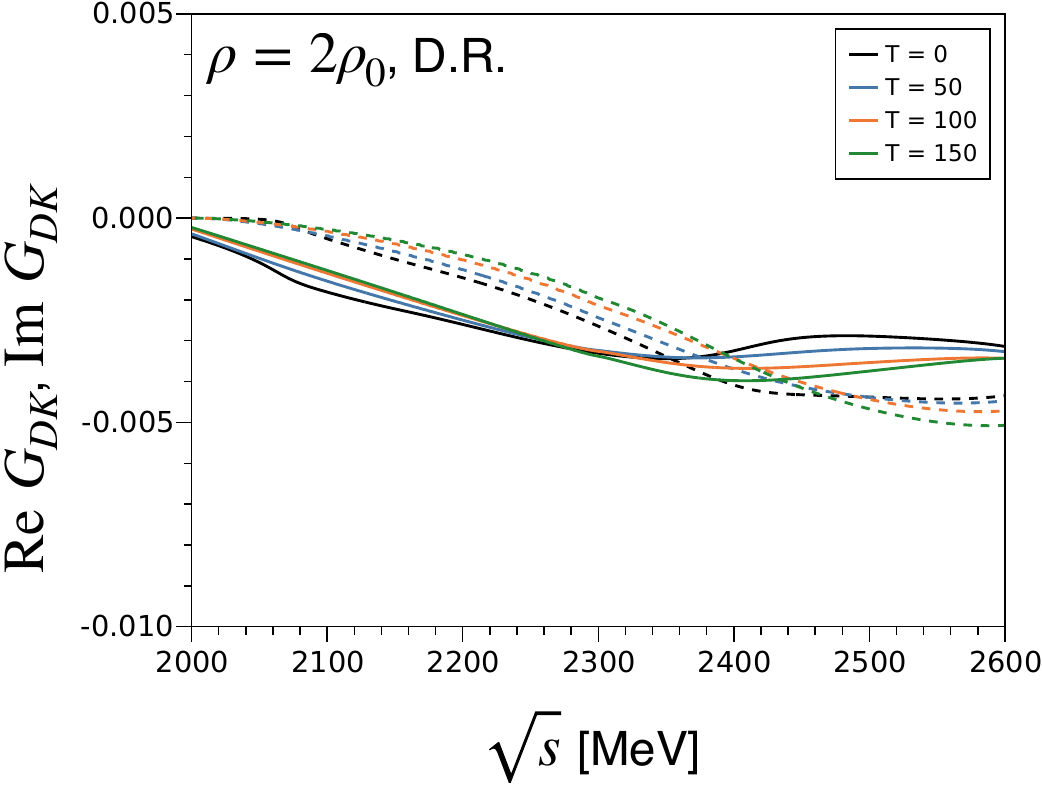}
    \caption{Real (solid lines) and imaginary (dashed lines) parts of the $DK$ loop function $G_{DK}$ in matter as functions of the center-of-mass energy $\sqrt{s}$, for the cutoff regularization scheme (upper panels) and the DR scheme (lower panels). The left panels in both regularization schemes are obtained at fixed $\rho = \rho_0$ for different temperatures ($T=0,50,100,150$ MeV), whereas the right panels correspond to the calculation at $2 \rho_0$ for the same temperatures.}
    \label{fig:GDK-med}
\end{figure*}

\begin{figure*}
    \centering
\includegraphics[width=0.45\linewidth]
{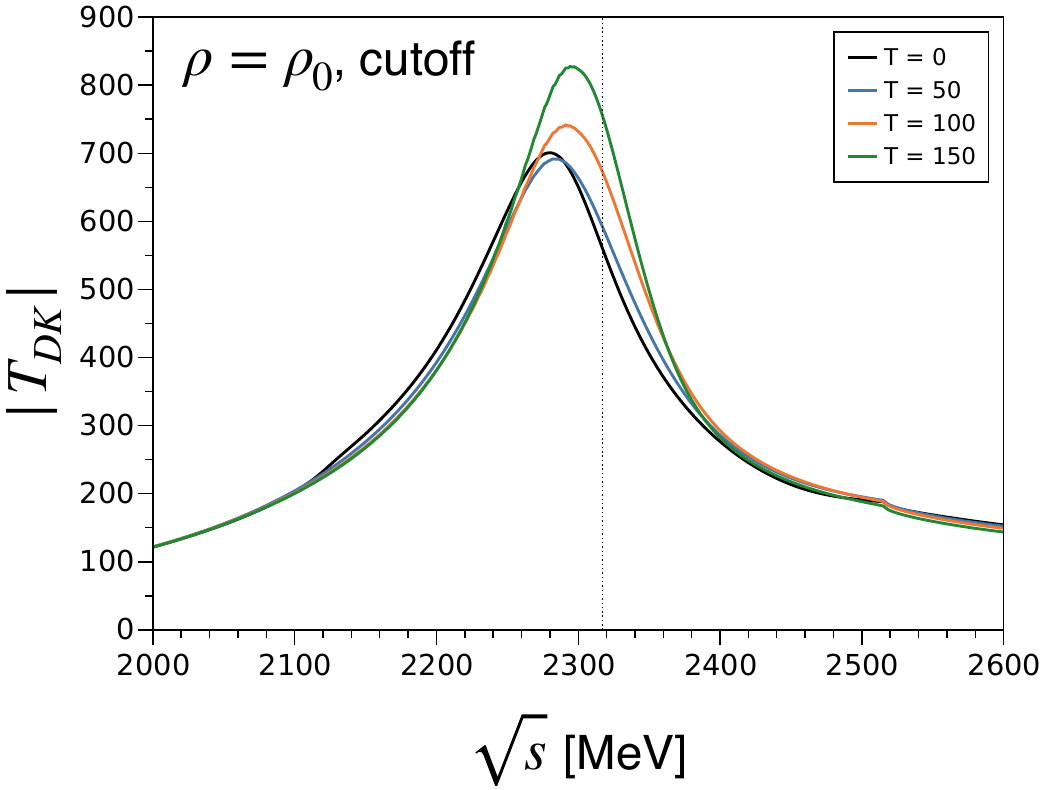}
\includegraphics[width=0.45\linewidth]
{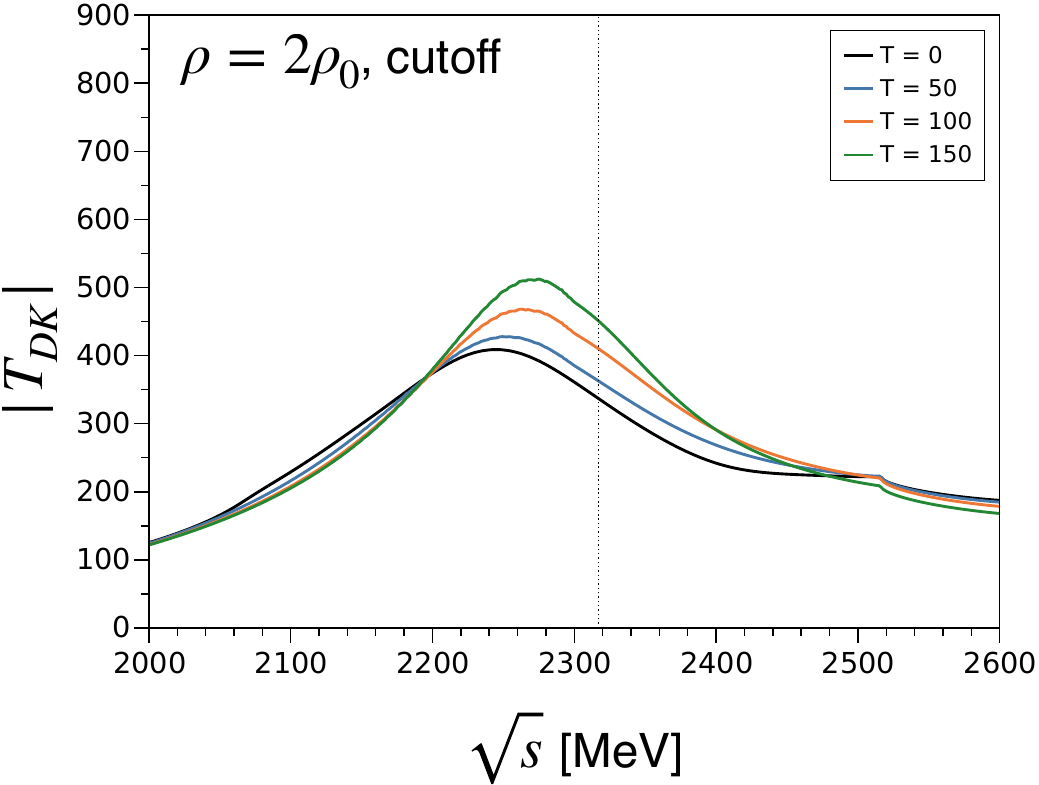}
\includegraphics[width=0.45\linewidth]
{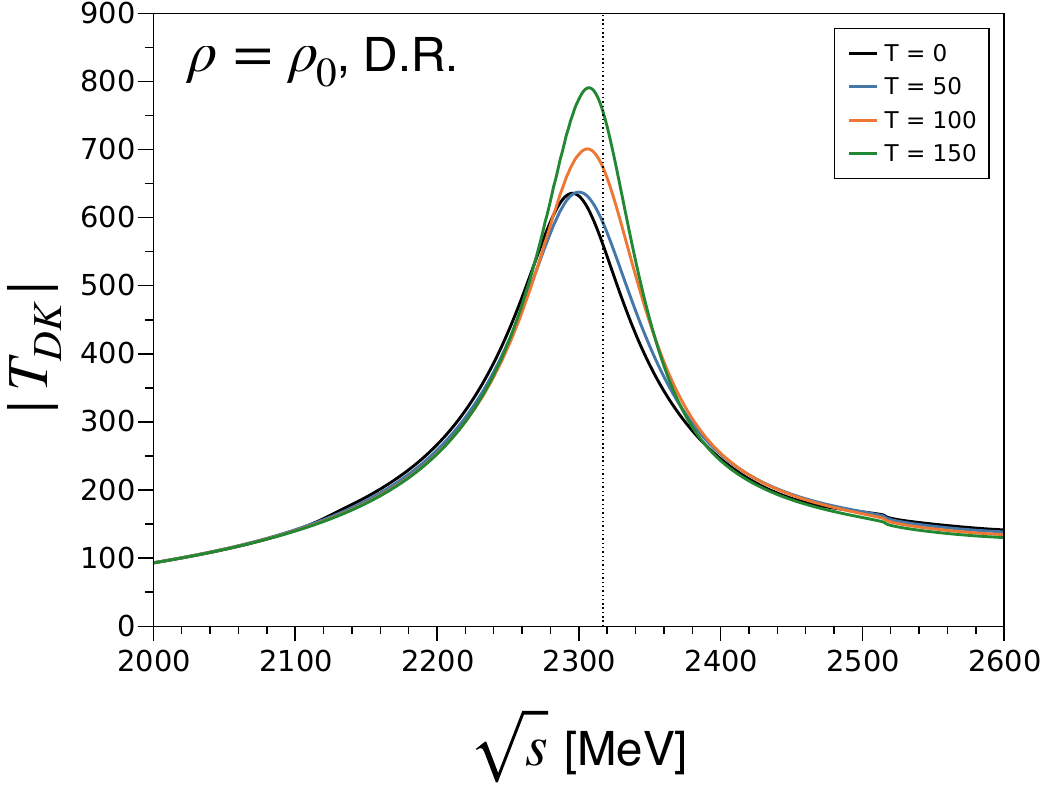}
\includegraphics[width=0.45\linewidth]
{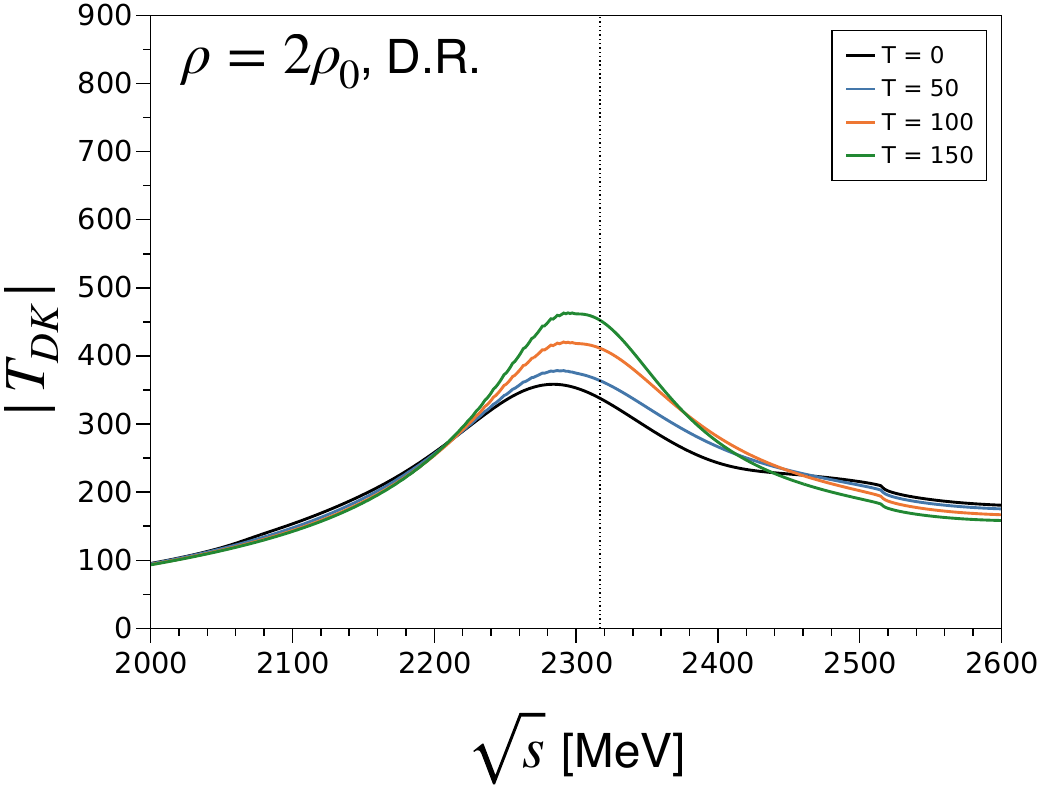}
    \caption{The modulus of the $DK$ scattering amplitude $|T_{DK}|$ in medium as a function of the center-of-mass energy $\sqrt{s}$ for $\rho = \rho_0$ (left panels) and $\rho = 2\rho_0$ (right panels) for the cutoff regularization scheme (upper panels) and the DR scheme (lower panels).}
    \label{fig:TDK-med}
\end{figure*}

\section{The $D_{s0}^*(2317)^\pm$ \\ in a hot and dense medium}
\label{sec:medium}

We start this section by analyzing the behavior of the $D_{s0}^*(2317)^+$ in a hot and dense medium.
As detailed in Sec.~\ref{sec:formalism}, the in-medium modifications are implemented such that the $DK$ loop explicitly incorporates both finite-density and finite-temperature effects, whereas the $D_s\eta$ loop is subjected only to temperature corrections. 

In Fig.~\ref{fig:GDK-med} we display the energy dependence of the $DK$ loop function for various temperatures and two distinct nuclear matter densities: normal saturation density ($\rho_0$, left panels) and twice saturation density ($2\rho_0$, right panels). The results are shown for both regularization methods, with the cutoff scheme in the top panels and the DR scheme in the bottom panels. Both regularization frameworks yield highly consistent results. Notably, the explicit temperature effects on the loop functions are moderate, whereas the density-driven modifications are significantly more pronounced. This difference is directly tied to the behavior of the underlying $D$-meson spectral function studied in Ref.~\cite{Tolos:2007vh}, which exhibits a much stronger sensitivity to density changes than to thermal variations.

The modulus of the in-medium $DK$ scattering amplitude is shown in Fig.~\ref{fig:TDK-med} across various temperatures for both $\rho_0$ (left panels) and $2\rho_0$ (right panels) and for the two regularization schemes (cutoff regularization in the upper plots and DR in the lower ones). Consistent with our findings for the loop functions, the amplitude exhibits a much stronger dependence on density than on temperature. 

As the density increases, the peak corresponding to the $D_{s0}^*(2317)^+$ shifts toward lower energies—moving further away from its free-space mass— and undergoes significant broadening. Conversely, when the temperature is raised at a fixed density, the peak shifts back toward the free-space $D_{s0}^*(2317)^+$ mass and becomes noticeably narrower. This narrowing at higher temperatures is counterintuitive at first glance, since one would conventionally expect thermal effects to induce additional collisional broadening, as was indeed found in the study of thermal medium effects on the $D$ mesons and their chiral partners of Ref.~\cite{Montana:2020lfi}.

To understand why the peak shifts to higher energies with temperature, we examine the behavior of the $D$-meson spectral function of Ref.~\cite{Tolos:2007vh} at zero momentum. It is found there that, due to the thermal smearing of the nuclear matter Fermi surface, the quasiparticle peak of the $D$ meson moves closer to its vacuum mass value. Physically, this occurs because the in-medium self-energy receives increasing contributions from higher-momentum $DN$ pairs that experience a weaker overall interaction.

The narrowing of the $D_{s0}^*(2317)^+$ state with temperature can also be traced back to the features of the $D$-meson self-energy detailed in Ref.~\cite{Tolos:2007vh}. At $T=0$, the quasiparticle peak of the $D$ meson is strongly mixed with a $\Sigma_c(2800)N^{-1}$ hole excitation, which produces a relatively broad peak in nuclear matter. However, as the temperature increases, this $\Sigma_c(2800)N^{-1}$ excitation mode smears out significantly. Consequently, the $D$-meson spectral function recovers a narrower lineshape with a distinct quasiparticle peak and an extended, flat tail strength. This behavior explains the corresponding reduction in the width of the $D_{s0}^*(2317)^+$ with increasing temperature.

A previous result on the $D_{s0}^*(2317)^+$ lineshape in nuclear matter has been reported in Ref.~\cite{Montesinos:2024uhq}, where a repulsive effect in the position of the $D_{s0}^*(2317)^+$ with density is found. At $\rho_0$ the resonance appears about 30~MeV higher in energy with respect to its nominal mass, for a molecular probability of $P_0=0.8$, which is close to the value associated to the HMChPT models as the one employed here. Moreover, the $D_{s0}^*(2317)^+$ is narrower, with a width of 40~MeV at $\rho_0$ in contrast to the width of 150~MeV that we obtain in the present calculation. These differences are mostly attributed to the different $D$-meson spectral function employed in the $DK$ loop.
The study of Ref.~\cite{Montesinos:2024uhq} implemented the 
$D$-meson spectral function obtained in Ref.~\cite{Tolos:2009nn}, which presents a distinct quasiparticle well separated from other resonant-hole excitation modes, $\Sigma_c(2823) N^{-1}$ and $\Sigma_c(2868) N^{-1}$, located higher up in energy. We note that these $\Sigma_c$ resonances are predictions of the model and are located at higher energies than the experimental $\Sigma_c(2800)$ state. As mentioned above, the $D$-meson spectral function used in the present work  mixes the quasiparticle peak with the $\Sigma_c(2800) N^{-1}$ mode, making it appear broader \cite{Tolos:2007vh}. This clearly explains the wider in-medium $D_{s0}^*(2317)^+$ found here with respect to that found in Ref.~\cite{Montesinos:2024uhq}. To understand the difference of almost 40 MeV towards higher energies in the position of the $D_{s0}^*(2317)^+$  compared to our results, we have performed various numerical tests and can attribute this shift in Ref.~\cite{Montesinos:2024uhq} to the accumulation of three similarly-sized repulsive effects: the different $D$-meson spectral function, the incorporation of a repulsive mass shift for the $K$-meson (not included in the present work), and the absence of coupling of the $DK$ channel to the $D_s \eta$ one.

\begin{figure*}
    \centering
\includegraphics[width=0.45\linewidth]
{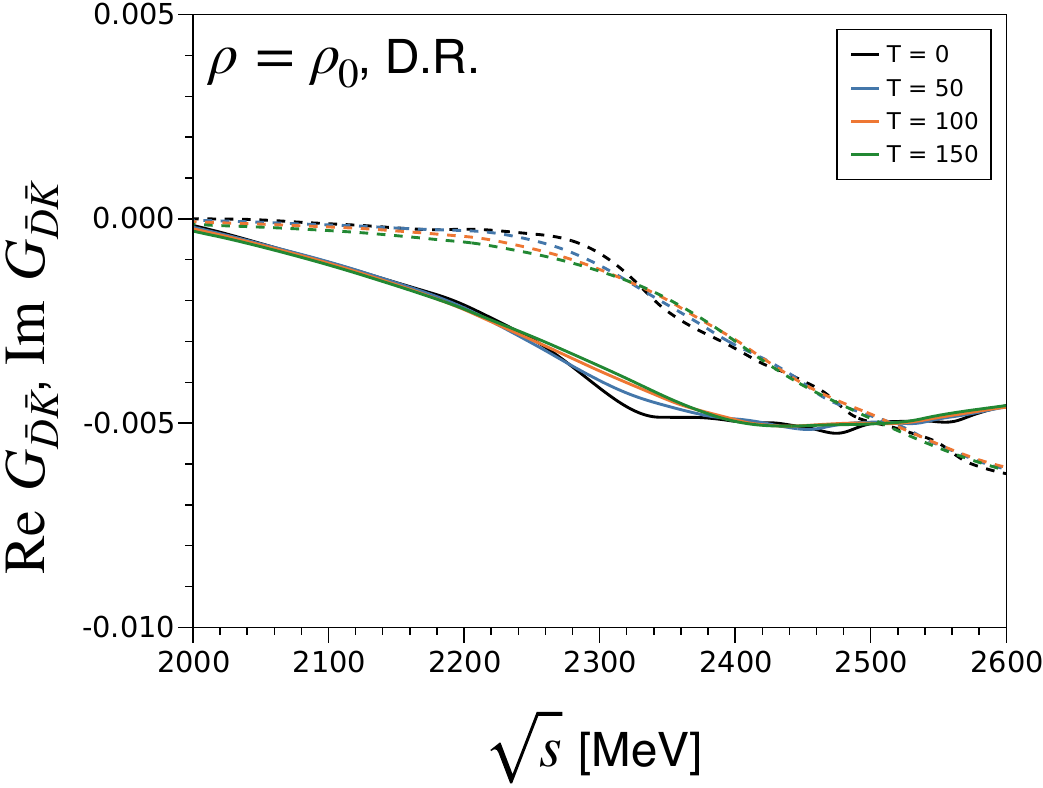}
\includegraphics[width=0.45\linewidth]
{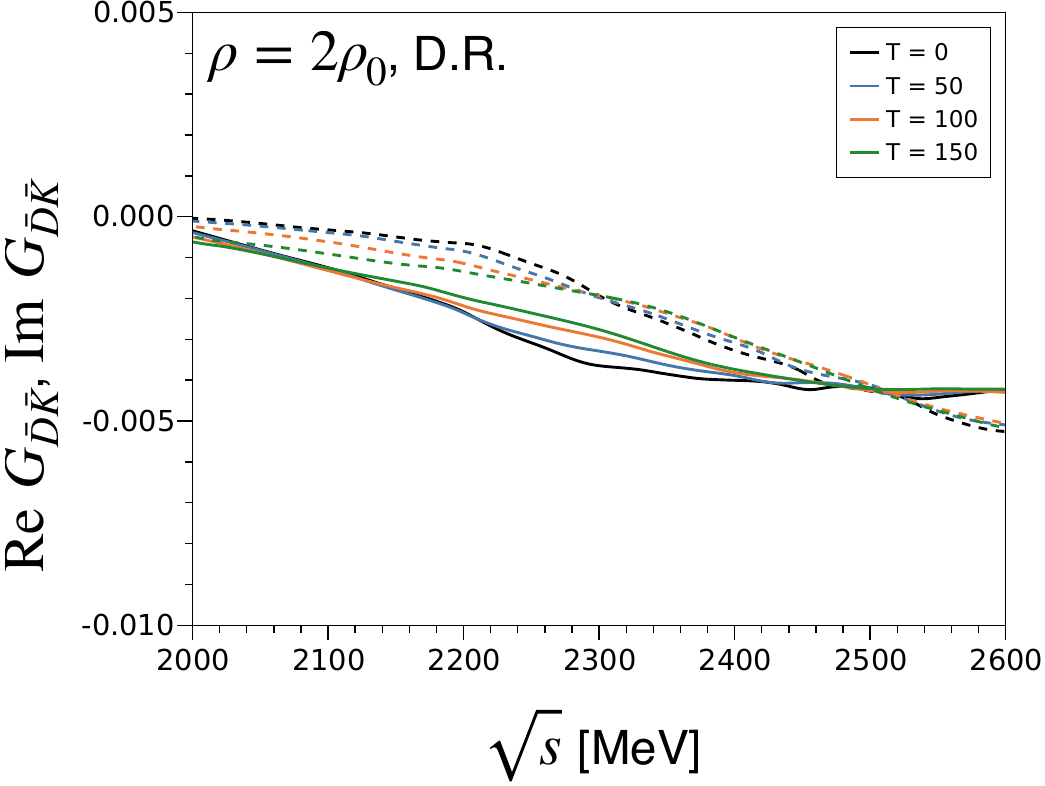}
    \caption{Real (solid lines) and imaginary (dashed lines) parts of the $\bar D \bar K$ loop function $G_{\bar{D}\bar{K}}$ in matter as functions of the center-of-mass energy $\sqrt{s}$ within the DR scheme. The left plot is obtained at fixed $\rho = \rho_0$ for different temperatures ($T=0,50,100,150$ MeV), whereas the right plot corresponds to the calculation at $2 \rho_0$ for the same temperatures.}
        \label{fig:GDbarKbar-med}
\end{figure*}

\begin{figure*}
    \centering
\includegraphics[width=0.45\linewidth]
{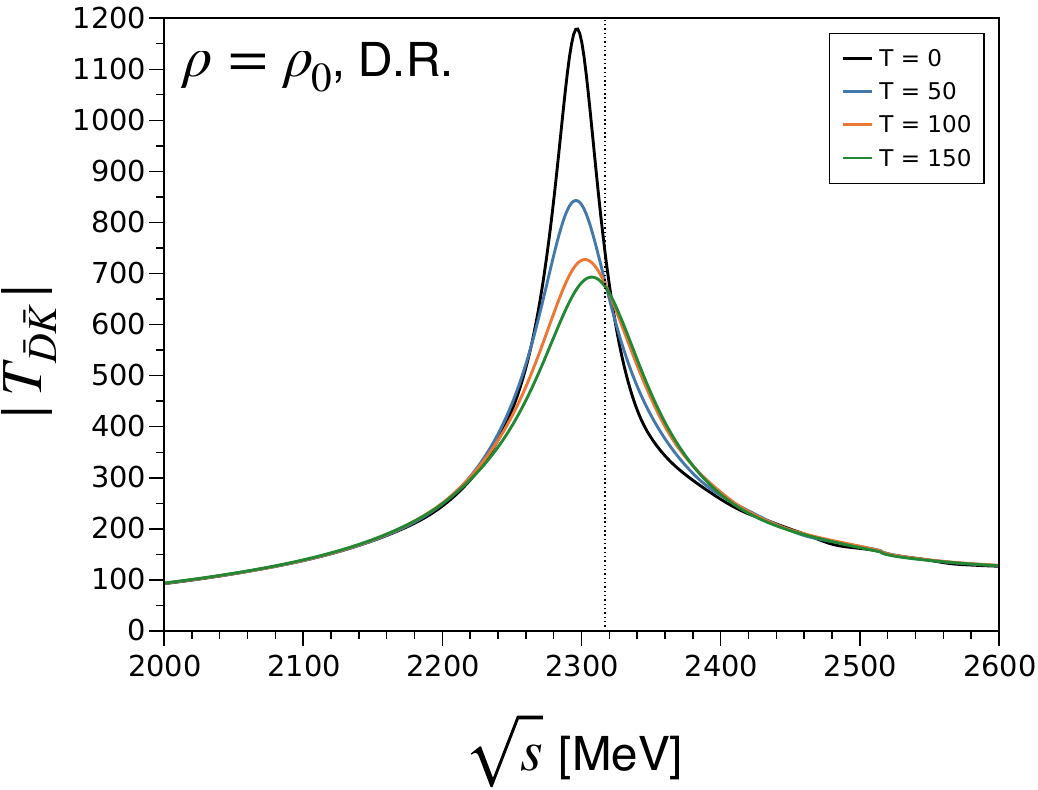}
\includegraphics[width=0.45\linewidth]
{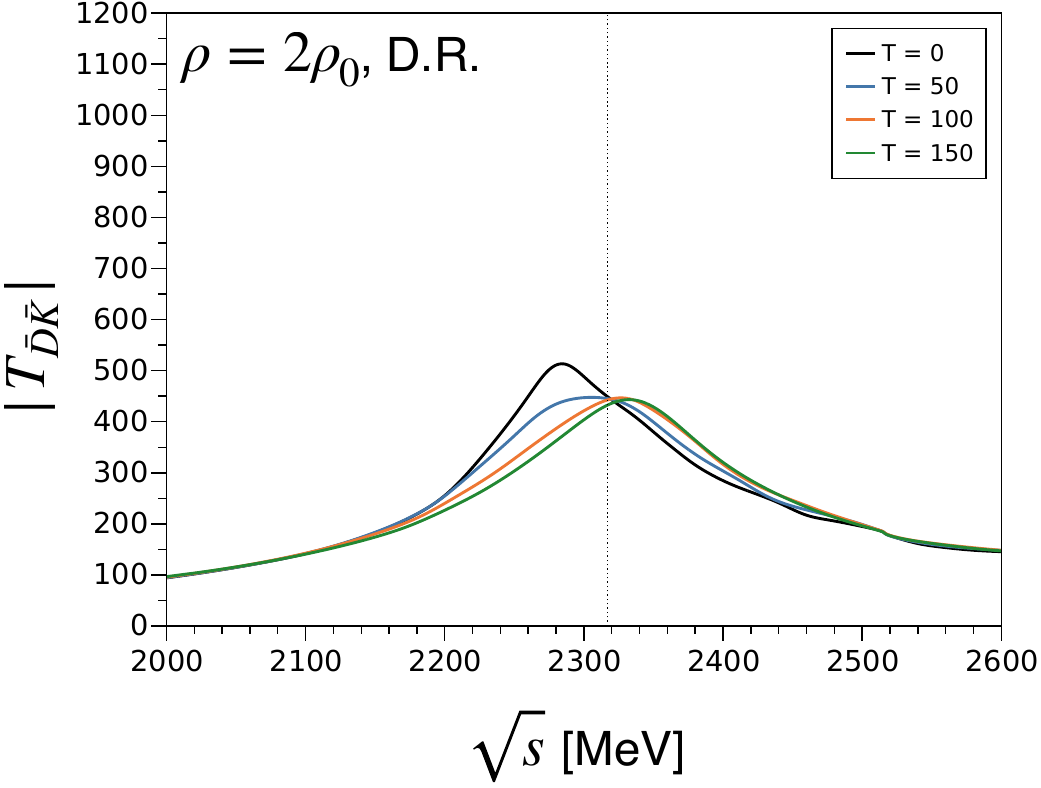}
    \caption{The modulus of the $\bar{D}\bar{K}$ scattering amplitude $|T_{\bar D \bar K}|$ in medium as a function of the center-of-mass energy $\sqrt{s}$ for $\rho = \rho_0$ (left panel) and $\rho = 2\rho_0$ (right panel) within the DR scheme.}
    \label{fig:TDbarKbar-DR}
\end{figure*}

As for the case of $D_{s0}^*(2317)^-$, we show in Fig.~\ref{fig:GDbarKbar-med} the energy dependence of the $\bar D \bar K$ loop function for various temperatures and both $\rho_0$ (left panel) and $2\rho_0$ (right panel) within the DR scheme. The corresponding modulus of the in-medium $\bar D \bar K$ scattering amplitude in that regularization scheme is displayed in Fig.~\ref{fig:TDbarKbar-DR} for $\rho_0$ (left panel) and $2\rho_0$ (right panel). 

At $\rho = \rho_0$ and $T=0$, the $D_{s0}^*(2317)^-$ peak is located roughly $15\text{ MeV}$ below its nominal free-space mass and acquires a width of about $60\text{ MeV}$. For non-zero temperatures, increasing the density causes the peak position of the $D_{s0}^*(2317)^-$ to shift slightly toward higher energies, while the state becomes significantly broader. Similar to the case of the $D_{s0}^*(2317)^+$, the in-medium $D_{s0}^*(2317)^-$ shifts closer to its free-space mass as the temperature increases. However, in contrast to the positive charge state, its width increases with temperature and eventually saturates at higher values of $T$. These modifications are driven by the behavior of the $\bar K$ spectral function. As discussed in \cite{Tolos:2008di}, the $\bar K$ spectral function features a prominent peak that widens at larger nuclear densities due to enhanced collision and absorption processes. With increasing temperature, this spectral function spans a broader range of energies, and its peak moves closer to the free-space position.

We note, however, that a direct comparison with the results for the $D_{s0}^*(2317)^-$ from Ref.~\cite{Montesinos:2024uhq} is not straightforward, as it requires a detailed analysis of the $\bar D \bar K$ loop in the nuclear medium. While both the previous computation in Ref.~\cite{Montesinos:2024uhq} and our present work incorporate the same $\bar K$ spectral function, they differ significantly in their treatment of the $\bar D$ meson. In our calculation, we neglect in-medium modifications to the $\bar D$ spectral function, given that only a minor mass increase with density is expected according to the model of Ref.~\cite{Tolos:2007vh}. Conversely, the study in Ref.~\cite{Montesinos:2024uhq} employs a $\bar D$ spectral function derived from the SU(8)-extended Weinberg-Tomozawa model of Ref.~\cite{Garcia-Recio:2008rjt}. As seen in Ref.~\cite{Montesinos:2023qbx}, that particular spectral function at $\rho_0$ exhibits a distinct quasiparticle peak located $30\text{ MeV}$ below the nominal $\bar D$ mass, alongside an additional structure associated to a pentaquark resonance generated by their specific $\bar D N$ interaction model.

Nevertheless, in agreement with the discussion in Ref.~\cite{Montesinos:2024uhq}, we observe a noticeably different in-medium pattern for the $D_{s0}^*(2317)^+$ compared to its antiparticle, the $D_{s0}^*(2317)^-$. This asymmetry stems from the distinct behaviors of their respective $DK$ and $\bar D \bar K$ components in hot dense matter, thereby offering a promising new avenue for elucidating the internal structure of these exotic states.

\begin{figure*}
    \centering
\includegraphics[width=0.45\linewidth]
{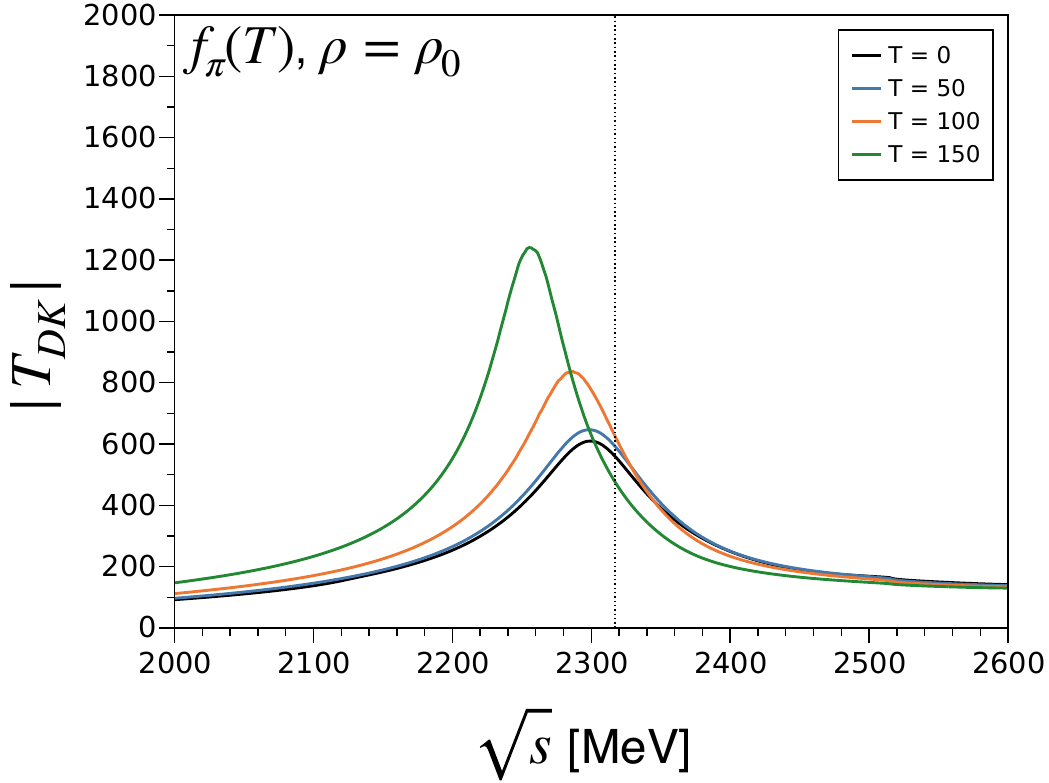}
\includegraphics[width=0.45\linewidth]
{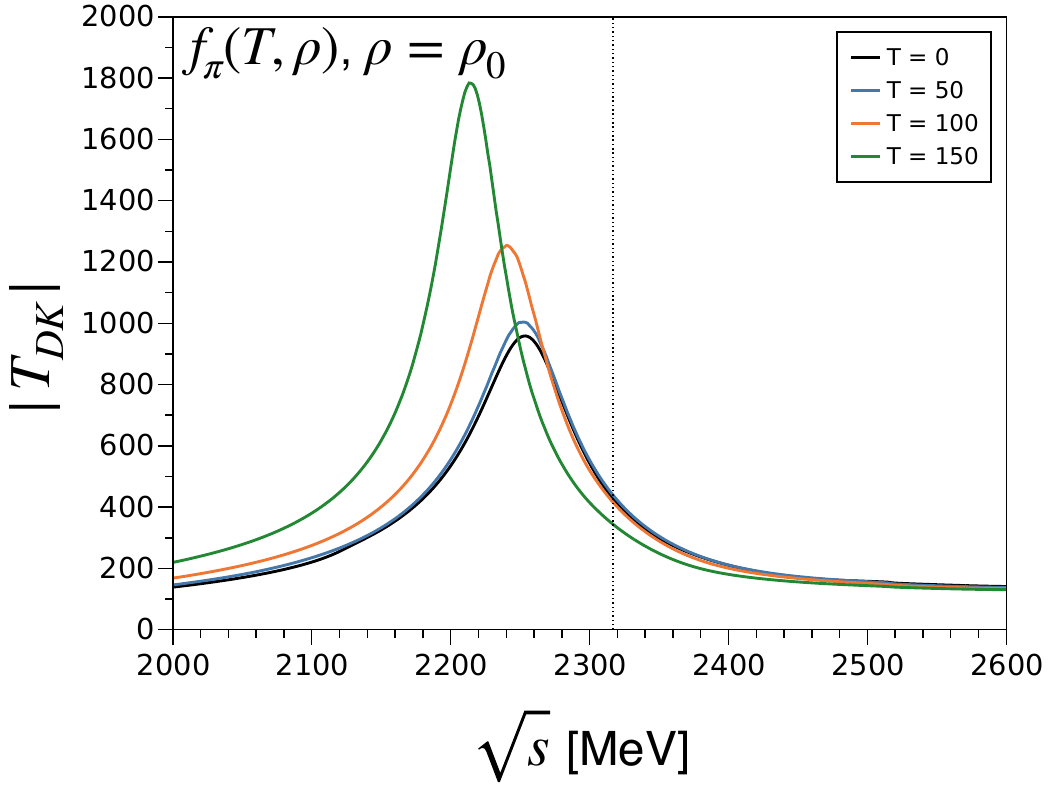}
    \caption{The modulus of the $DK$ scattering amplitude $|T_{DK}|$ in medium as a function of the center-of-mass energy $\sqrt{s}$, using $f_\pi(T)$ (left panel) and $f_\pi(T,\rho)$ (right panel) for $\rho = \rho_0$ and various temperatures.}
    \label{fig:TDK_f(T)_f(T,rho)_rho0}
\end{figure*}

\section{Effects of Medium-Dependent Interactions}
\label{sec:medium_dependent}

Next, we analyze the impact of incorporating explicit temperature and density dependencies into the interaction kernel itself, driven by the in-medium changes of the pion decay constant $f_{\pi}(T, \rho)$ as expressed in Eq.~(\ref{eq:f_pi_T_rho}). 

The left panel of Fig.~\ref{fig:TDK_f(T)_f(T,rho)_rho0} presents the $DK$ amplitude for $\rho_0$ when keeping only the temperature-dependent correction factor. Because the value of $f_{\pi}(T)$ decreases with temperature, the effective strength of the attractive $DK$ Weinberg-Tomozawa interaction increases, leading to a more strongly bound $D_{s0}^*(2317)^+$ state. When the full density-dependent term is additionally implemented, we obtain the results displayed in the right panel of Fig.~\ref{fig:TDK_f(T)_f(T,rho)_rho0}. Here, the $D_{s0}^*(2317)^+$ state shifts even further down in energy due to the additional attraction induced by density. Interestingly, as the $D_{s0}^*(2317)^+$ state becomes lighter, its decay width diminishes. This is directly related to the suppression of absorption processes in the lower-energy tail of the $D$-meson spectral function, which manifests as a net decrease in the imaginary part of the $DK$ loop at lower energies.

\begin{figure*}
    \centering
\includegraphics[width=0.45\linewidth]
{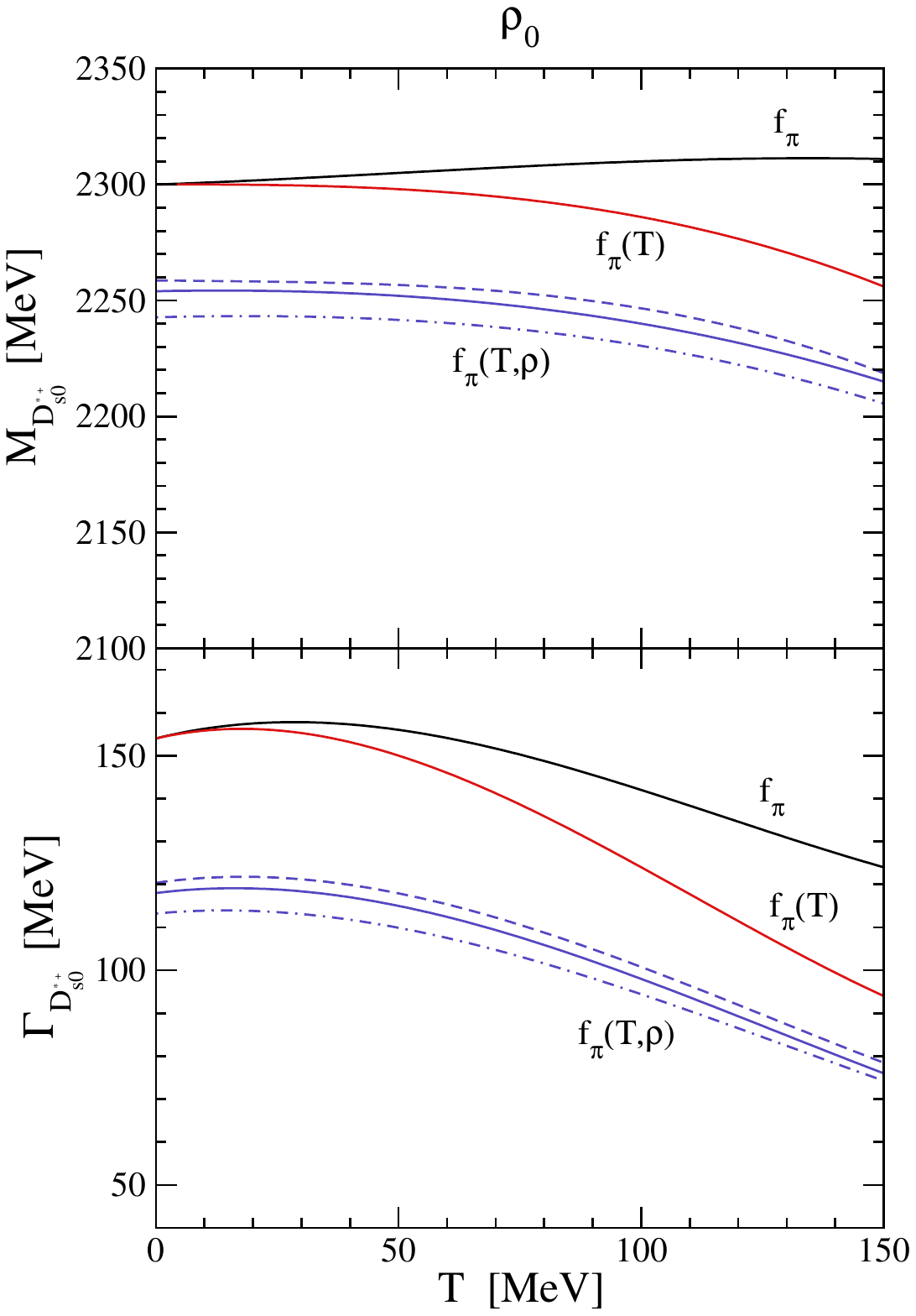}
\includegraphics[width=0.45\linewidth]
{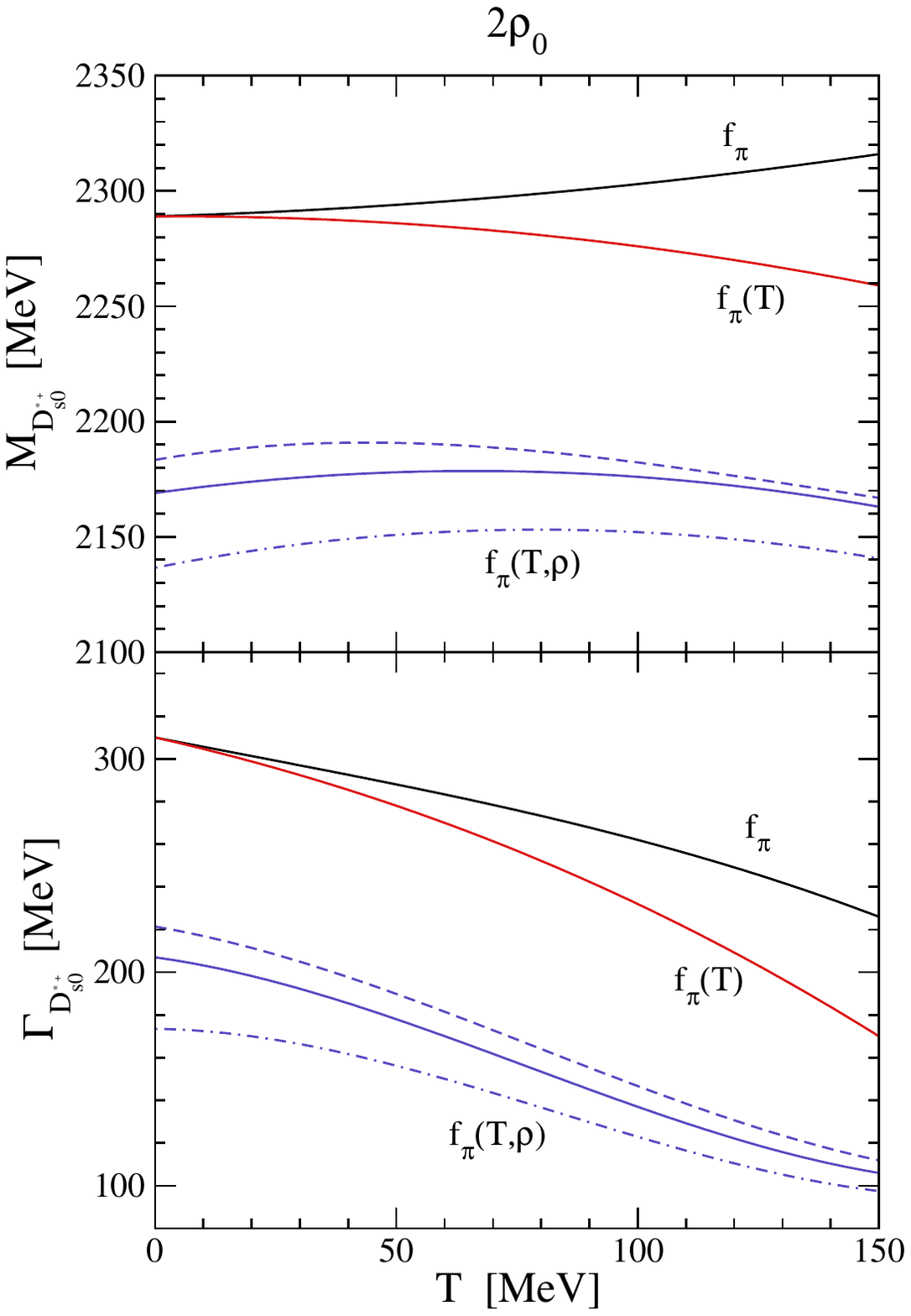}
    \caption{Mass (top panels) and width (bottom panels) of the $D_{s0}^*(2317)^+$ in nuclear matter at $\rho=\rho_0$ (left panels) and $\rho=2\rho_0$ (right panels) as functions of temperature, for different $f_\pi$ prescriptions. The dashed, solid, and dot-dashed blue lines are obtained using $\sigma_{\pi N}=45$, 50 and 60 MeV, respectively.}
    \label{fig:Ds_rho0_2rho0}
\end{figure*}

Our complete results for the mass and width of the $D_{s0}^*(2317)^+$ as a function of temperature are summarized in Fig.~\ref{fig:Ds_rho0_2rho0} for densities of $\rho_0$ (left panels) and $2\rho_0$ (right panels). We compare three distinct physical scenarios for the coupling: a medium-independent decay constant $f_{\pi}$ (black lines), a purely temperature-dependent $f_{\pi}(T)$ (red lines), and the full medium-dependent $f_{\pi}(T, \rho)$ case (blue lines). The solid blue line is obtained employing $\sigma_{\pi N}=50$ MeV, chosen as our reference value for the pion-nucleon sigma term. The dashed and dash-dotted blue lines are obtained using $\sigma_{\pi N}=45$ and 60 MeV, respectively, to cover the range of values found in the literature. 

When the interaction kernel is kept medium-independent, the mass of the $D_{s0}^*(2317)^+$ slightly increases with temperature while its width narrows. This thermal narrowing is further enhanced when $f_{\pi}$ is systematically reduced by both temperature and density effects. Most notably, the mass of the in-medium $D_{s0}^*(2317)^+$ drops sharply when the density-dependent modification of $f_{\pi}$ is incorporated: at zero temperature and saturation density ($\rho_0$), the mass drops by nearly 50~MeV from the value near 2300~MeV obtained in the case of a medium-independent decay constant. This drop is amplified to roughly 120~MeV at twice saturation density ($2\rho_0$). 

By comparison, the mass shifts directly driven by temperature effects are much more moderate. For instance, at $\rho_0$, the medium must be heated to approximately $T = 150$~MeV to achieve a 50~MeV mass drop comparable to the one driven purely by the density dependence of $f_{\pi}$. At $2\rho_0$, the temperature dependence of $f_{\pi}$ has an even slighter effect on the $T=0$ baseline mass value. In short, across all three interaction scenarios explored, the mass of the $D_{s0}^*(2317)^+$ remains significantly less sensitive to thermal changes than to density variations.

Similarly to the previous analysis for $D_{s0}^*(2317)^+$,  in Fig.~\ref{fig:TDbarKbar_f(T)_f(T,rho)_rho0} we also show the $\bar D \bar K$ amplitude for $\rho_0$ when only temperature corrections are included in the pion decay constant (left panel) together with the $\bar D \bar K$ amplitude when temperature and density corrections are simultaneously incorporated (right panel). As in the case of $D_{s0}^*(2317)^+$, the decrease in the value of $f_{\pi}(T)$ leads to a more attractive $\bar D \bar K$ interaction that results in a more bound $D_{s0}^*(2317)^-$ state. When not only temperature but also density modifications are incorporated, the $D_{s0}^*(2317)^-$ shifts further down since the $\bar D \bar K$ interaction becomes more attractive, as already observed for $D_{s0}^*(2317)^+$. However, contrary to the case of $D_{s0}^*(2317)^+$, the decay width increases until saturation when temperature corrections are implemented whereas it substantially increases when both temperature and density corrections are considered. This result is a direct consequence of the behavior of the $\bar K$ spectral function in matter. The $\bar K$ quasiparticle peak widens with density and temperature due to enhanced collision and absorption processes, as previously discussed in Fig.~\ref{fig:TDbarKbar-DR}. 

\begin{figure*}
    \centering
\includegraphics[width=0.45\linewidth]
{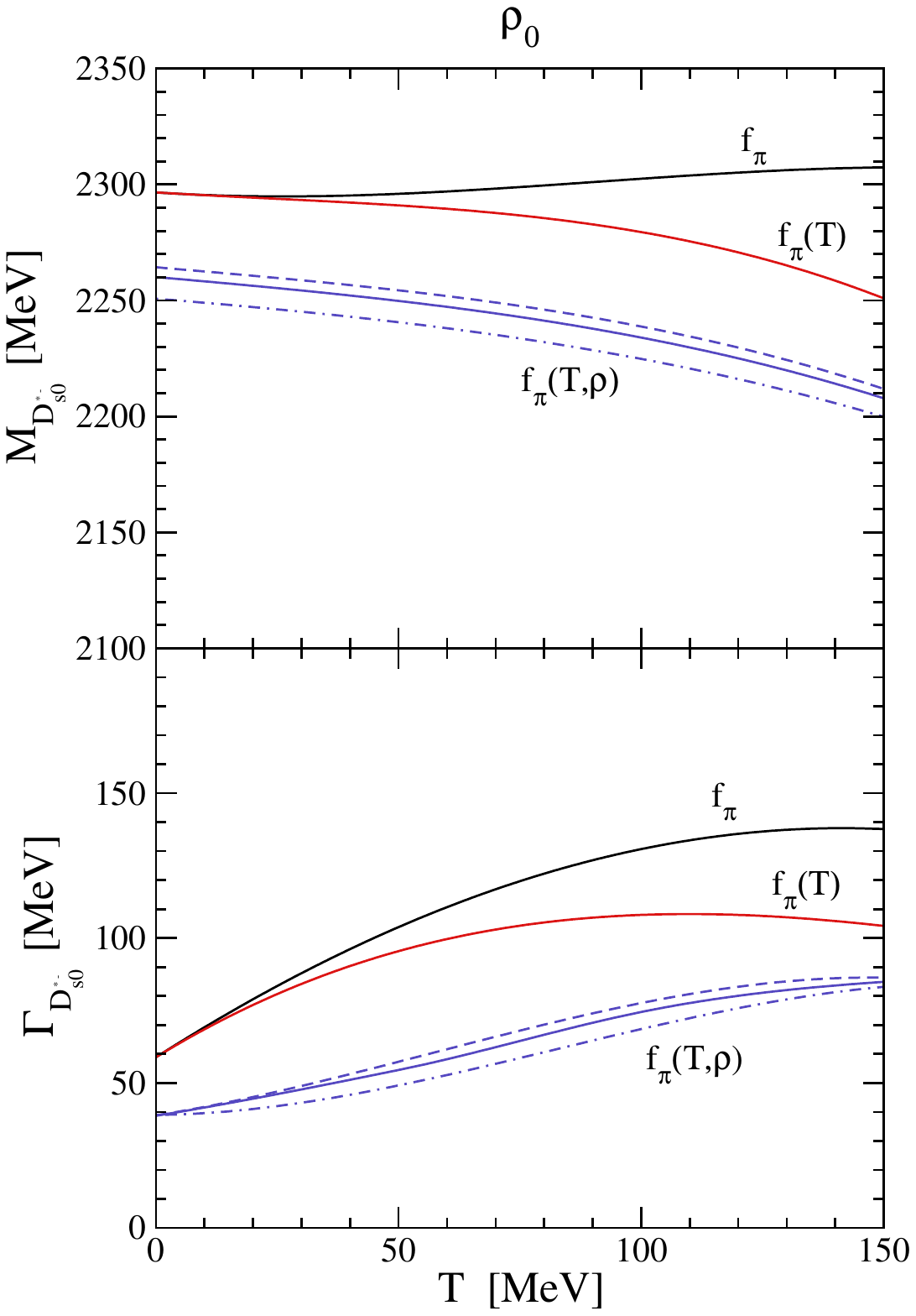}
\includegraphics[width=0.45\linewidth]
{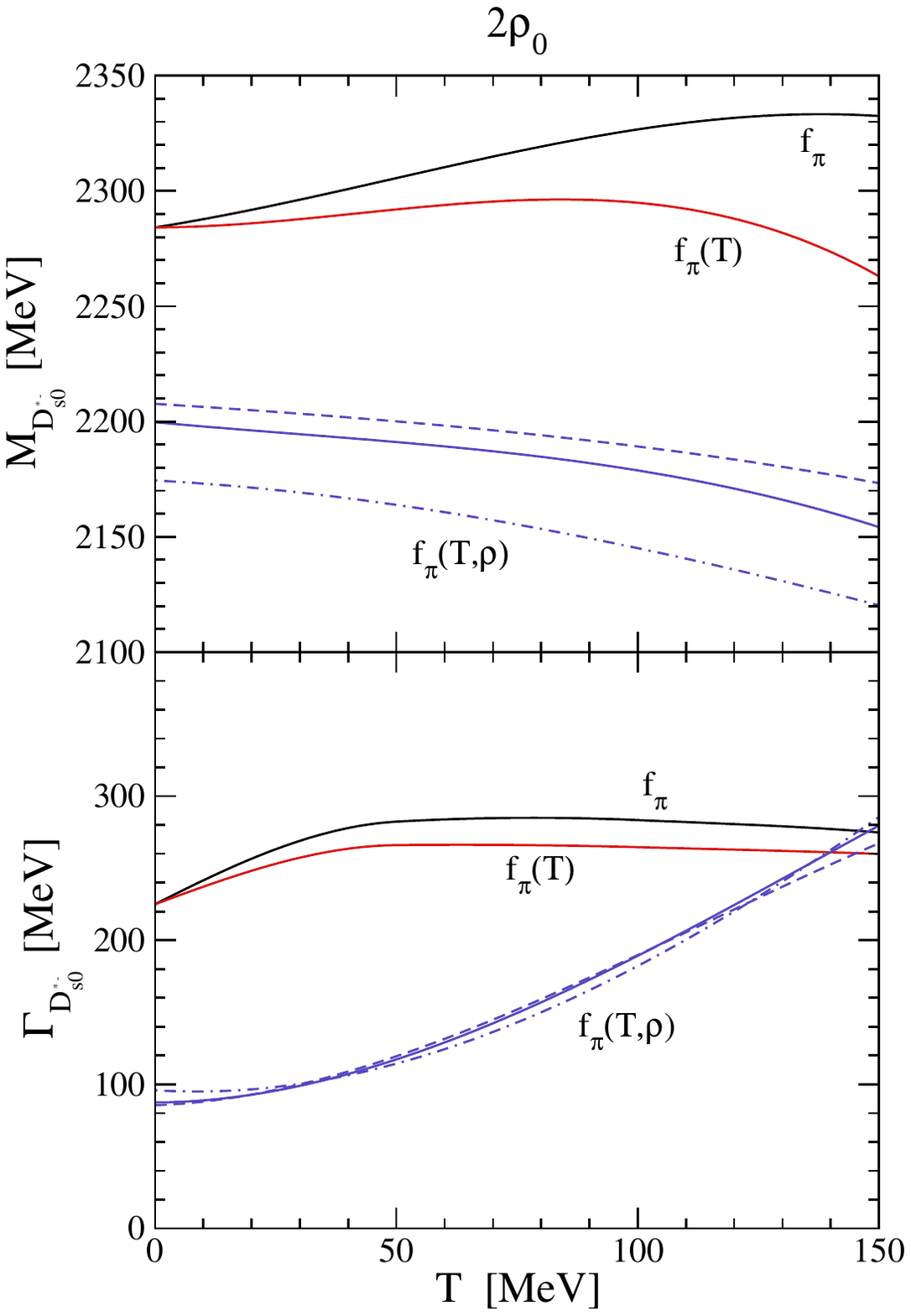}
    \caption{Mass (top panels) and width (bottom panels) of the $D_{s0}^*(2317)^-$ in nuclear matter at $\rho=\rho_0$ (left panels) and $\rho=2\rho_0$ (right panels) as functions of temperature, for different $f_\pi$ prescriptions. The dashed, solid, and dot-dashed blue lines are obtained using $\sigma_{\pi N}=45$, 50 and 60 MeV, respectively.}
    \label{fig:Dsminus_rho0_2rho0}
\end{figure*}

The full results for the mass and width of $D_{s0}^*(2317)^-$ are displayed in Fig.~\ref{fig:Dsminus_rho0_2rho0} as a function of temperature for $\rho_0$ (left panels) and $2\rho_0$ (right panels) for the three different $f_{\pi}$ prescriptions. When the interaction is independent of the medium, the mass of the $D_{s0}^*(2317)^-$ slightly increases with temperature while its width increases until saturating around 150 MeV. In the case of the medium-modified kernel, the $D_{s0}^*(2317)^-$ mass tends to drop with temperature. Overall, the 
$D_{s0}^*(2317)^-$ mass stays less sensitive to
thermal changes than to density variations, as already observed for $D_{s0}^*(2317)^+$. 
As for the  width, it increases until saturating when the interaction is kept medium independent or when only temperature corrections are incorporated. In the case of including temperature and density modifications in the pion decay constant, the width increases with temperature, especially at $2 \rho_0$, when it changes by $\sim$200 MeV from low to high temperatures. This is again tied to the widening of the ${\bar K}$ spectral function at this density and, especially, to the extended low-energy tail that it develops \cite{Tolos:2008di}. In particular, the insensitivity observed in the width to the employed value of $\sigma_{\pi N}$ is precisely due to the constant and structureless ${\bar K}$ strength at the lower energies explored by a less massive $D_{s0}^*(2317)^-$.

We note that the inclusion of a medium-dependent pion decay constant $f_\pi(\rho, T)$ in the interaction kernel introduces a formal inconsistency, as the same medium modifications are not simultaneously implemented in the calculation of the $D$ and $\bar{K}$ spectral functions. Therefore, the results discussed in this section are exploratory in nature, aimed at providing a qualitative assessment of how kernel modifications might alter the properties of the $D_{s0}^*(2317)^+$ and of its antiparticle. The fully consistent treatment remains a task for future work.

\begin{figure*}
    \centering
\includegraphics[width=0.45\linewidth]
{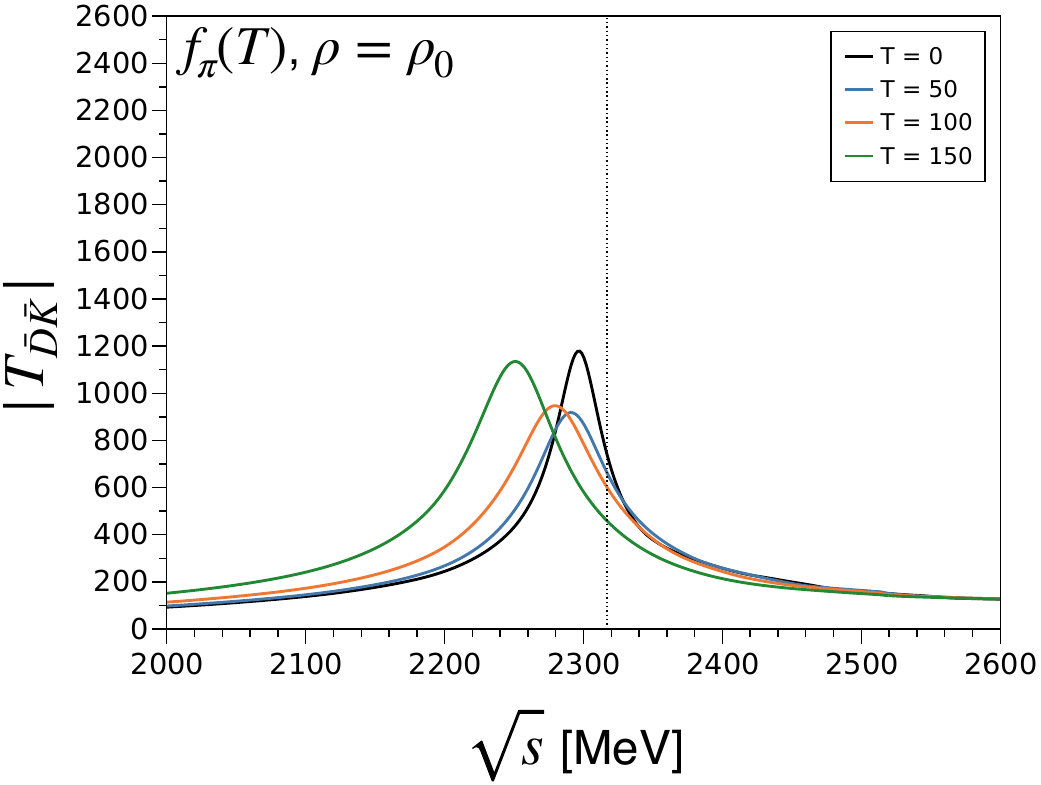}
\includegraphics[width=0.45\linewidth]
{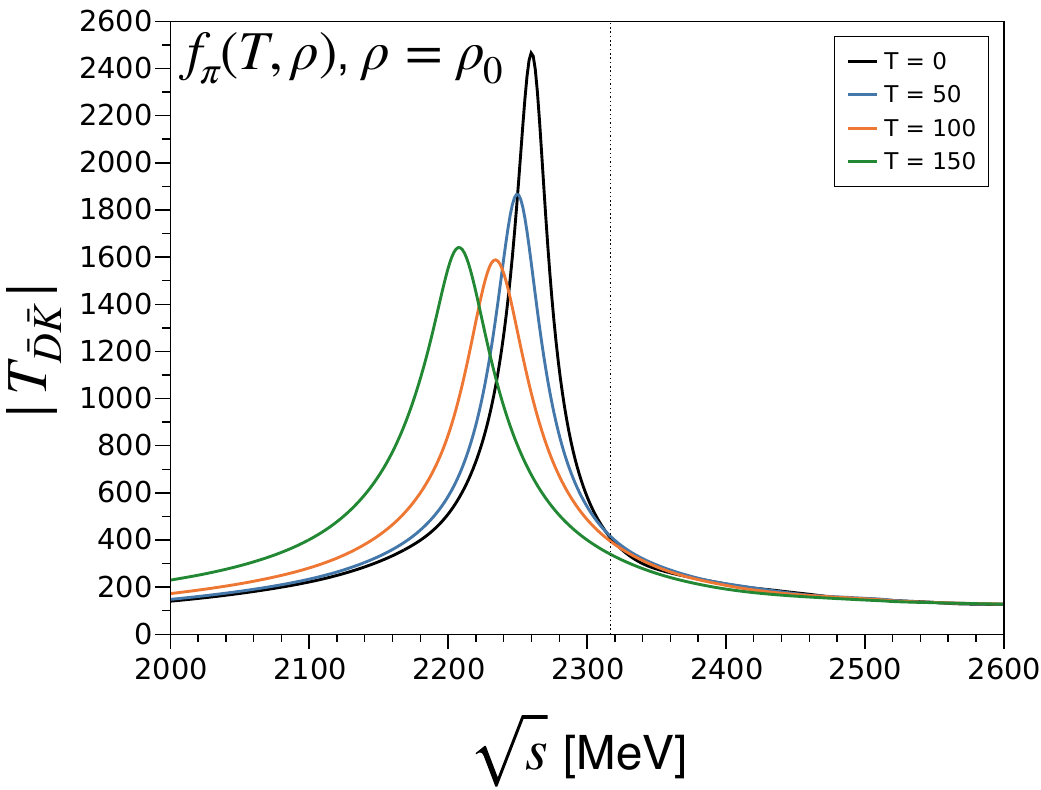}
    \caption{The modulus of the $\bar{D}\bar{K}$ scattering amplitude $|T_{\bar{D}\bar{K}}|$ in medium as a function of the center-of-mass energy $\sqrt{s}$, using $f_\pi(T)$ (left panel) and $f_\pi(T,\rho)$ (right panel) for $\rho = \rho_0$ and various temperatures.}
    \label{fig:TDbarKbar_f(T)_f(T,rho)_rho0}
\end{figure*}

\section{Conclusions} \label{sec:conclusions}
In this work we study the properties of $D_{s0}^*(2317)^\pm$ in hot and dense matter for conditions of density and/or temperature produced in HICs at RHIC, LHC or future CBM/FAIR energies.
 The $D_{s0}^*(2317)^+$ and $D_{s0}^*(2317)^-$ are generated dynamically by the $DK$ and $\bar D \bar K$ interactions (and associated $D_s \eta$ and $\bar D_s \eta$ channels), respectively, obtained from a HMChPT model at NLO.

We find that nuclear density drives the most pronounced changes, whereas thermal effects are less important. As density increases, the $D_{s0}^*(2317)^+$ peak shifts lower in energy below its vacuum mass and broadens significantly, tracking the behavior of the $D$-meson spectral function. Raising the temperature counters this density shift as the thermal smearing of the nuclear matter Fermi surface effectively reduces the strength of the interaction shifting the $D_{s0}^*(2317)^+$ peak back toward its vacuum mass. In addition, the melting of the $\Sigma_c(2800)N^{-1}$ hole excitation in the $D$-meson spectral function causes the $D_{s0}^*(2317)^+$ peak to narrow with temperature.
Discrepancies with the results of Ref.~\cite{Montesinos:2024uhq}, which found a narrower width and a repulsive mass shift, are traced to differences in the underlying $D$-meson spectral function, the inclusion of a repulsive $K$-meson shift in the previous publication, and the consideration of the $DK$-$D_s\eta$ coupled channels in the present work.

As for the antiparticle $D_{s0}^*(2317)^-$, this state is governed by the in-medium $\bar{K}$ spectral function. Similarly to the positively charged partner, the mass of the $D_{s0}^*(2317)^-$ decreases when the density of the medium is increased. While temperature similarly pulls its mass back toward the vacuum value, its width broadens and saturates at high temperatures instead of narrowing. 

We also analyze the effect of including explicit density and temperature dependence into the interaction kernel via the in-medium pion decay constant $f_\pi(T, \rho)$. This medium-modified interaction strengthens the attractive $DK$ and $\bar{D}\bar{K}$ interactions, leading to a substantial mass drop of the respective $D_{s0}^*(2317)^+$ and $D_{s0}^*(2317)^-$ dynamically generated states. This effect increases with temperature and, at $T=150$ MeV, the mass of the $D_{s0}^*(2317)^\pm$ can be lowered by almost 100 MeV at $\rho_0$ and by almost 150~MeV at $2\rho_0$.
As for the width, the reduced value of the in-medium pion decay constant $f_\pi(T, \rho)$ further narrows the lineshape of the $D_{s0}^*(2317)^+$ due to the suppression of absorption processes in the lower-energy tail of the $D$-meson spectral function.
Conversely, the decay width of the $D_{s0}^*(2317)^-$ increases significantly when both temperature and density modifications are included. This substantial broadening is driven directly by the behavior of the $\bar{K}$ spectral function, which widens in the medium due to enhanced collision and absorption processes.

In summary, the asymmetry in how the $D_{s0}^*(2317)^+$ and  $D_{s0}^*(2317)^- $states behave in matter could help to constrain the internal structure of these exotic states in experimental setups probing hot dense matter, such as the CBM experiment at FAIR.

\begin{appendix}

\section{Low-energy constants and numerical inputs}
\label{sec:LECs}

Here, we summarize the numerical inputs used in this work. The low-energy constants in the effective Lagrangian are listed in Table~\ref{tab:LECs}. The isospin coefficients required to construct the interaction kernel are given in Table~\ref{tab:coefficients}. In Table~\ref{tab:parameters}, we also provide the parameters associated with the regularization of the loop functions, including the sharp cutoff $\Lambda$ and the subtraction constant $a(\mu)$ at the regularization scale $\mu$, as well as the integration ranges ($\omega_{\rm min}$ and $\omega_{\rm max}$) employed in the numerical evaluation of the in-medium loop functions.

\begin{table}
 \caption{The low-energy constants for $D\Phi$ scattering in the effective Lagrangian~\eqref{eq:NLO} taken from Fit-2B in Ref.~\cite{Guo:2018tjx}. \label{tab:LECs}}
 \begin{ruledtabular}
  \begin{tabular}{cccccc}
    $h_{0}$ & $h_{1}$ & $h_{2}$ & $h_{3}$ & $h_{4}$ [MeV$^{-2}$] & $h_{5}$ [MeV$^{-2}$] \\ \hline 
     $0.033$ & $0.45$ & $-0.12$ & $1.67$ & $-0.0054\times 10^{-6}$ & $-0.22\times 10^{-6}$ \\
  \end{tabular}
  \end{ruledtabular}
\end{table}

\begin{table*}
 \caption{The isospin coefficients in the interaction kernel in Eq.~\eqref{eq:V}. $m_{\pi}$, $m_{K}$, and $m_{\eta}$ stand for the masses of $\pi$, $K$, and $\eta$ mesons, respectively.\label{tab:coefficients}}
 \begin{ruledtabular}
  \begin{tabular}{c|ccccc}
    Channel ($i \to j$) & $C^{\rm LO}_{ij}$ & $C^{0}_{ij}$ & $C^{1}_{ij}$ & $C^{24}_{ij}$ & $C^{35}_{ij}$  \\ \hline 
     $DK \to DK$ & -2 & $m_{K}^{2}$ & $-2m_{K}^{2}$ & 1 & 2 \\
     $DK \to D_{s}\eta$ & $-\sqrt{3}$ & 0 & $-\frac{1}{2\sqrt{3}}(5m_K^2-3m_\pi^2)$ & 0 & $1/\sqrt{3}$\\
     $D_{s}\eta \to D_{s}\eta$ & 0 & $m_{\eta}^{2}$ & $-\frac{4}{3}(2m_K^2-m_\pi^2)$ & 1 & $4/3$ \\
  \end{tabular}
  \end{ruledtabular}
\end{table*}

\begin{table*}
 \caption{Numerical parameters used for the loop functions.\label{tab:parameters}}
 \begin{ruledtabular}
  \begin{tabular}{ccccccc}
    $\Lambda$ [MeV] & $a(\mu)$ & $\mu$ [MeV] & $\omega_{\rm min}$ for $DK$ [MeV] & $\omega_{\rm max}$ for $DK$ [MeV] & $\omega_{\rm min}$ for $\bar{D}\bar{K}$ [MeV] & $\omega_{\rm max}$ for $\bar{D}\bar{K}$ [MeV] \\ \hline 
     615 & -1.79 & 1000 & 1510 & 2450 & 50 & 1500 \\
  \end{tabular}
  \end{ruledtabular}
\end{table*}

\section{Analytic expressions of the loop function in free space}
\label{sec:loop-function-free}

Here, we summarize the expressions of the regularized two-meson loop function in free space in Eq.~\eqref{eq:G-free}. We first give the expression obtained in the cutoff regularization scheme, where a finite cutoff $\Lambda$ is introduced in the three-momentum integral~\cite{Oller:1998hw,Montana:2020vjg}:
\begin{widetext}
\begin{align}
G_{l}^{\rm cut}(s) &= \frac{1}{32\pi^{2}}\left\{\log\left(\frac{m_{D}^{2}m_{\Phi}^{2}}{\Lambda^{4}}\right) - \frac{m_{D}^{2} - m_{\Phi}^{2}}{s}\log\left(\frac{m_{\Phi}^{2}}{m_{D}^{2}}\right) \right. \nonumber \\
&\quad + 2\frac{m_{D}^{2} - m_{\Phi}^{2}}{s}\log\left(\frac{1 + \sqrt{1 + m_{\Phi}^{2}\Lambda^{-2}}}{1 + \sqrt{1 + m_{D}^{2}\Lambda^{-2}}}\right) - 2\log\left[\left(1 + \sqrt{1 + \frac{m_{\Phi}^{2}}{\Lambda^{2}}}\right)\left(1 + \sqrt{1 + \frac{m_{D}^{2}}{\Lambda^{2}}}\right)\right] \nonumber \\
& \quad + \frac{2 q_{l}}{\sqrt{s}}\left[\log\left(s - (m_{D}^{2} - m_{\Phi}^{2}) + 2\sqrt{s}q_{l}\sqrt{1 + m_{\Phi}^{2}\Lambda^{-2}}\right) + \log\left(s + (m_{D}^{2} - m_{\Phi}^{2}) + 2\sqrt{s}q_{l}\sqrt{1 + m_{D}^{2}\Lambda^{-2}}\right) \right. \nonumber \\
& \quad \left. \left. - \log\left(-s + (m_{D}^{2} - m_{\Phi}^{2}) + 2\sqrt{s}q_{l}\sqrt{1 + m_{\Phi}^{2}\Lambda^{-2}}\right) - \log\left(-s - (m_{D}^{2} - m_{\Phi}^{2}) + 2\sqrt{s}q_{l}\sqrt{1 + m_{D}^{2}\Lambda^{-2}}\right) \right] \right \}.
\end{align}
\end{widetext}
Here, $q_{l}$ denotes the relative momentum between the two mesons in the $l$-th channel propagating in the loop:
\begin{align}
q_{l} = \frac{\sqrt{[s - (m_{D} + m_{\Phi})^{2}][s - (m_{D} - m_{\Phi})^{2}]}}{2\sqrt{s}}.
\end{align}

As another regularization scheme, we consider DR, in which the two-meson loop function is given by~\cite{Guo:2018tjx,Montana:2020vjg}
\begin{widetext}
\begin{align}
G_{l}^{\rm DR}(s) &= \frac{1}{16\pi^{2}}\left\{a_{l}(\mu) + \log\frac{m_{D}^{2}}{\mu^2} + \frac{m_{\Phi}^{2} - m_{D}^{2} + s}{2s} \log\frac{m_{\Phi}^{2}}{m_{D}^{2}}\right.\nonumber \\
&\quad + \frac{q_{l}}{\sqrt{s}}\left[\right. \log(s - (m_{D}^{2} - m_{\Phi}^{2}) + 2q_{l}\sqrt{s}) + \log(s + (m_{D}^{2} - m_{\Phi}^{2}) + 2q_{l}\sqrt{s}) \nonumber \\
&\quad \left.\left. -\log(-s + (m_{D}^{2} - m_{\Phi}^{2}) + 2q_{l}\sqrt{s}) - \log(-s - (m_{D}^{2} - m_{\Phi}^{2}) + 2q_{l}\sqrt{s}) \right]\right\}.
\end{align}
\end{widetext}
Here, $a_{l}(\mu)$ is the channel-dependent subtraction constant at the regularization scale $\mu$.

\end{appendix}

\begin{acknowledgments}
The authors would like to thank Makoto Oka, Philipp Gubler, and José Manuel Alarcón Soriano for useful discussions. This work is supported by the grants CEX2020-001058-M and CEX2024-001451-M of the Spanish program ``Unidad de Excelencia María de Maeztu", financed by MCIN/AEI/10.13039/501100011033, the MaX-CSIC Excellence Award MaX4-SOMMA-ICE, the projects PID2023-147112NB-C21, PID2022-139427NB-I00 and PID2025-168786NB-I00 financed by the Spanish MCIN/AEI/10.13039/501100011033 ERDF/EU, by the Grant CIPROM 2023/59 of Generalitat Valenciana, and by JSPS KAKENHI Grant Number JP25K23387. T.K. is supported by the RIKEN special postdoctoral researcher program.
\end{acknowledgments}

\bibliography{references}

@article{ParticleDataGroup:2024cfk,
    author = "Navas, S. and others",
    collaboration = "Particle Data Group",
    title = "{Review of particle physics}",
    doi = "10.1103/PhysRevD.110.030001",
    journal = "Phys. Rev. D",
    volume = "110",
    number = "3",
    pages = "030001",
    year = "2024"
}

@article{BaBar:2003oey,
    author = "Aubert, B. and others",
    collaboration = "BaBar",
    title = "{Observation of a narrow meson decaying to $D_s^+ \pi^0$ at a mass of 2.32-GeV/c$^2$}",
    eprint = "hep-ex/0304021",
    archivePrefix = "arXiv",
    reportNumber = "SLAC-PUB-9711, BABAR-PUB-03-011",
    doi = "10.1103/PhysRevLett.90.242001",
    journal = "Phys. Rev. Lett.",
    volume = "90",
    pages = "242001",
    year = "2003"
}

@article{CLEO:2003ggt,
    author = "Besson, D. and others",
    collaboration = "CLEO",
    title = "{Observation of a narrow resonance of mass 2.46-GeV/c**2 decaying to D*+(s) pi0 and confirmation of the D*(sJ)(2317) state}",
    eprint = "hep-ex/0305100",
    archivePrefix = "arXiv",
    reportNumber = "CLNS-03-1826, CLEO-03-09, CLNS03-1826",
    doi = "10.1103/PhysRevD.68.032002",
    journal = "Phys. Rev. D",
    volume = "68",
    pages = "032002",
    year = "2003",
    note = "[Erratum: Phys.Rev.D 75, 119908 (2007)]"
}

@article{Belle:2003kup,
    author = "Mikami, Y. and others",
    collaboration = "Belle",
    title = "{Measurements of the $D_{sJ}$ resonance properties}",
    eprint = "hep-ex/0307052",
    archivePrefix = "arXiv",
    doi = "10.1103/PhysRevLett.92.012002",
    journal = "Phys. Rev. Lett.",
    volume = "92",
    pages = "012002",
    year = "2004"
}

@article{Godfrey:1985xj,
    author = "Godfrey, S. and Isgur, Nathan",
    title = "{Mesons in a Relativized Quark Model with Chromodynamics}",
    doi = "10.1103/PhysRevD.32.189",
    journal = "Phys. Rev. D",
    volume = "32",
    pages = "189--231",
    year = "1985"
}

@article{Godfrey:1986wj,
    author = "Godfrey, Stephen and Kokoski, Richard",
    title = "{The Properties of p Wave Mesons with One Heavy Quark}",
    reportNumber = "TRI-PP-86-51, GIPP-90-6A",
    doi = "10.1103/PhysRevD.43.1679",
    journal = "Phys. Rev. D",
    volume = "43",
    pages = "1679--1687",
    year = "1991"
}

@article{Zeng:1994vj,
    author = "Zeng, J. and Van Orden, J. W. and Roberts, W.",
    title = "{Heavy mesons in a relativistic model}",
    eprint = "hep-ph/9412269",
    archivePrefix = "arXiv",
    reportNumber = "CEBAF-TH-94-08",
    doi = "10.1103/PhysRevD.52.5229",
    journal = "Phys. Rev. D",
    volume = "52",
    pages = "5229--5241",
    year = "1995"
}

@article{Gupta:1994mw,
    author = "Gupta, Suraj N. and Johnson, James M.",
    title = "{Quantum chromodynamic potential model for light heavy quarkonia and the heavy quark effective theory}",
    eprint = "hep-ph/9409432",
    archivePrefix = "arXiv",
    doi = "10.1103/PhysRevD.51.168",
    journal = "Phys. Rev. D",
    volume = "51",
    pages = "168--175",
    year = "1995"
}

@article{Lahde:1999ih,
    author = "Lahde, T. A. and Nyfalt, C. J. and Riska, D. O.",
    title = "{Spectra and M1 decay widths of heavy light mesons}",
    eprint = "hep-ph/9908485",
    archivePrefix = "arXiv",
    doi = "10.1016/S0375-9474(00)00154-8",
    journal = "Nucl. Phys. A",
    volume = "674",
    pages = "141--167",
    year = "2000"
}

@article{DiPierro:2001dwf,
    author = "Di Pierro, Massimo and Eichten, Estia",
    title = "{Excited Heavy - Light Systems and Hadronic Transitions}",
    eprint = "hep-ph/0104208",
    archivePrefix = "arXiv",
    reportNumber = "FERMILAB-PUB-01-033-T",
    doi = "10.1103/PhysRevD.64.114004",
    journal = "Phys. Rev. D",
    volume = "64",
    pages = "114004",
    year = "2001"
}

@article{Ebert:1997nk,
    author = "Ebert, D. and Galkin, V. O. and Faustov, R. N.",
    title = "{Mass spectrum of orbitally and radially excited heavy - light mesons in the relativistic quark model}",
    eprint = "hep-ph/9712318",
    archivePrefix = "arXiv",
    reportNumber = "HUB-EP-97-90",
    doi = "10.1103/PhysRevD.59.019902",
    journal = "Phys. Rev. D",
    volume = "57",
    pages = "5663--5669",
    year = "1998",
    note = "[Erratum: Phys.Rev.D 59, 019902 (1999)]"
}

@article{Alexandrou:2019tmk,
    author = "Alexandrou, Constantia and Berlin, Joshua and Finkenrath, Jacob and Leontiou, Theodoros and Wagner, Marc",
    title = "{Tetraquark interpolating fields in a lattice QCD investigation of the $D_{s0}^\ast(2317)$ meson}",
    eprint = "1911.08435",
    archivePrefix = "arXiv",
    primaryClass = "hep-lat",
    doi = "10.1103/PhysRevD.101.034502",
    journal = "Phys. Rev. D",
    volume = "101",
    number = "3",
    pages = "034502",
    year = "2020"
}

@article{Yang:2021tvc,
    author = "Yang, Zhi and Wang, Guang-Juan and Wu, Jia-Jun and Oka, Makoto and Zhu, Shi-Lin",
    title = "{Novel Coupled Channel Framework Connecting the Quark Model and Lattice QCD for the Near-threshold Ds States}",
    eprint = "2107.04860",
    archivePrefix = "arXiv",
    primaryClass = "hep-ph",
    doi = "10.1103/PhysRevLett.128.112001",
    journal = "Phys. Rev. Lett.",
    volume = "128",
    number = "11",
    pages = "112001",
    year = "2022"
}

@article{Colangelo:2003vg,
    author = "Colangelo, P. and De Fazio, F.",
    title = "{Understanding D(sJ)(2317)}",
    eprint = "hep-ph/0305140",
    archivePrefix = "arXiv",
    reportNumber = "BARI-TH-03-462",
    doi = "10.1016/j.physletb.2003.08.003",
    journal = "Phys. Lett. B",
    volume = "570",
    pages = "180--184",
    year = "2003"
}

@article{Dai:2003yg,
    author = "Dai, Yuan-Ben and Huang, Chao-Shang and Liu, Chun and Zhu, Shi-Lin",
    title = "{Understanding the D+(sJ)(2317) and D+(sJ)(2460) with sum rules in HQET}",
    eprint = "hep-ph/0306274",
    archivePrefix = "arXiv",
    doi = "10.1103/PhysRevD.68.114011",
    journal = "Phys. Rev. D",
    volume = "68",
    pages = "114011",
    year = "2003"
}

@article{Narison:2003td,
    author = "Narison, Stephan",
    title = "{Open charm and beauty chiral multiplets in QCD}",
    eprint = "hep-ph/0307248",
    archivePrefix = "arXiv",
    doi = "10.1016/j.physletb.2004.11.002",
    journal = "Phys. Lett. B",
    volume = "605",
    pages = "319--325",
    year = "2005"
}

@article{Bardeen:2003kt,
    author = "Bardeen, William A. and Eichten, Estia J. and Hill, Christopher T.",
    title = "{Chiral multiplets of heavy - light mesons}",
    eprint = "hep-ph/0305049",
    archivePrefix = "arXiv",
    reportNumber = "FERMILAB-PUB-03-071-T",
    doi = "10.1103/PhysRevD.68.054024",
    journal = "Phys. Rev. D",
    volume = "68",
    pages = "054024",
    year = "2003"
}

@article{Lee:2004gt,
    author = "Lee, Ian Woo and Lee, Taekoon and Min, D. P. and Park, Byung-Yoon",
    title = "{Chiral radiative corrections and D(s)(2317)/D(2308) mass puzzle}",
    eprint = "hep-ph/0412210",
    archivePrefix = "arXiv",
    doi = "10.1140/epjc/s10052-006-0149-7",
    journal = "Eur. Phys. J. C",
    volume = "49",
    pages = "737--741",
    year = "2007"
}

@article{Wang:2006bs,
    author = "Wang, Z. G. and Wan, S. L.",
    title = "{Structure of the D(s0)(2317) and the strong coupling constant g(D(s0)) DK with the light-cone QCD sum rules}",
    eprint = "hep-ph/0603007",
    archivePrefix = "arXiv",
    doi = "10.1103/PhysRevD.73.094020",
    journal = "Phys. Rev. D",
    volume = "73",
    pages = "094020",
    year = "2006"
}

@article{Cheng:2003kg,
    author = "Cheng, Hai-Yang and Hou, Wei-Shu",
    title = "{B decays as spectroscope for charmed four quark states}",
    eprint = "hep-ph/0305038",
    archivePrefix = "arXiv",
    doi = "10.1016/S0370-2693(03)00834-7",
    journal = "Phys. Lett. B",
    volume = "566",
    pages = "193--200",
    year = "2003"
}

@article{Terasaki:2003qa,
    author = "Terasaki, K.",
    title = "{BABAR resonance as a new window of hadron physics}",
    eprint = "hep-ph/0305213",
    archivePrefix = "arXiv",
    reportNumber = "YITP-03-28",
    doi = "10.1103/PhysRevD.68.011501",
    journal = "Phys. Rev. D",
    volume = "68",
    pages = "011501",
    year = "2003"
}

@article{Chen:2004dy,
    author = "Chen, Yu-Qi and Li, Xue-Qian",
    title = "{A Comprehensive four-quark interpretation of D(s)(2317), D(s)(2457) and D(s)(2632)}",
    eprint = "hep-ph/0407062",
    archivePrefix = "arXiv",
    doi = "10.1103/PhysRevLett.93.232001",
    journal = "Phys. Rev. Lett.",
    volume = "93",
    pages = "232001",
    year = "2004"
}

@article{Maiani:2004vq,
    author = "Maiani, L. and Piccinini, F. and Polosa, A. D. and Riquer, V.",
    title = "{Diquark-antidiquarks with hidden or open charm and the nature of X(3872)}",
    eprint = "hep-ph/0412098",
    archivePrefix = "arXiv",
    reportNumber = "ROMA1-1396-2004, FNT-T-2004-20, BA-TH-502-04, CERN-PH-TH-2004-239",
    doi = "10.1103/PhysRevD.71.014028",
    journal = "Phys. Rev. D",
    volume = "71",
    pages = "014028",
    year = "2005"
}

@article{Bracco:2005kt,
    author = "Bracco, M. E. and Lozea, A. and Matheus, Ricardo D'Elia and Navarra, F. S. and Nielsen, M.",
    title = "{Disentangling two- and four-quark state pictures of the charmed scalar mesons}",
    eprint = "hep-ph/0503137",
    archivePrefix = "arXiv",
    doi = "10.1016/j.physletb.2005.08.037",
    journal = "Phys. Lett. B",
    volume = "624",
    pages = "217--222",
    year = "2005"
}

@article{Wang:2006uba,
    author = "Wang, Zhi-Gang and Wan, Shao-Long",
    title = "{D(s)(2317) as a tetraquark state with QCD sum rules in heavy quark limit}",
    eprint = "hep-ph/0602080",
    archivePrefix = "arXiv",
    doi = "10.1016/j.nuclphysa.2006.07.041",
    journal = "Nucl. Phys. A",
    volume = "778",
    pages = "22--29",
    year = "2006"
}

@article{Browder:2003fk,
    author = "Browder, Thomas E. and Pakvasa, Sandip and Petrov, Alexey A.",
    title = "{Comment on the new D(s)(*)+ pi0 resonances}",
    eprint = "hep-ph/0307054",
    archivePrefix = "arXiv",
    reportNumber = "WSU-HEP-0306",
    doi = "10.1016/j.physletb.2003.10.067",
    journal = "Phys. Lett. B",
    volume = "578",
    pages = "365--368",
    year = "2004"
}

@article{vanBeveren:2003kd,
    author = "van Beveren, Eef and Rupp, George",
    title = "{Observed $D_s(2317)$ and tentative $D(2100\text{--}2300)$ as the
charmed cousins of the light scalar nonet}",
    eprint = "hep-ph/0305035",
    archivePrefix = "arXiv",
    doi = "10.1103/PhysRevLett.91.012003",
    journal = "Phys. Rev. Lett.",
    volume = "91",
    pages = "012003",
    year = "2003"
}

@article{Barnes:2003dj,
    author = "Barnes, T. and Close, F. E. and Lipkin, H. J.",
    title = "{Implications of a DK molecule at 2.32-GeV}",
    eprint = "hep-ph/0305025",
    archivePrefix = "arXiv",
    doi = "10.1103/PhysRevD.68.054006",
    journal = "Phys. Rev. D",
    volume = "68",
    pages = "054006",
    year = "2003"
}

@article{Szczepaniak:2003vy,
    author = "Szczepaniak, Adam P.",
    title = "{Description of the D*(s)(2320) resonance as the D pi atom}",
    eprint = "hep-ph/0305060",
    archivePrefix = "arXiv",
    doi = "10.1016/S0370-2693(03)00865-7",
    journal = "Phys. Lett. B",
    volume = "567",
    pages = "23--26",
    year = "2003"
}

@article{Kolomeitsev:2003ac,
    author = "Kolomeitsev, E. E. and Lutz, M. F. M.",
    title = "{On Heavy light meson resonances and chiral symmetry}",
    eprint = "hep-ph/0307133",
    archivePrefix = "arXiv",
    reportNumber = "GSI-PREPRINT-2003-20",
    doi = "10.1016/j.physletb.2003.10.118",
    journal = "Phys. Lett. B",
    volume = "582",
    pages = "39--48",
    year = "2004"
}

@article{Hofmann:2003je,
    author = "Hofmann, J. and Lutz, M. F. M.",
    title = "{Open charm meson resonances with negative strangeness}",
    eprint = "hep-ph/0308263",
    archivePrefix = "arXiv",
    reportNumber = "GSI-PREPRINT-2003-29",
    doi = "10.1016/j.nuclphysa.2003.12.013",
    journal = "Nucl. Phys. A",
    volume = "733",
    pages = "142--152",
    year = "2004"
}

@article{Guo:2006fu,
    author = "Guo, Feng-Kun and Shen, Peng-Nian and Chiang, Huan-Ching and Ping, Rong-Gang and Zou, Bing-Song",
    title = "{Dynamically generated 0+ heavy mesons in a heavy chiral unitary approach}",
    eprint = "hep-ph/0603072",
    archivePrefix = "arXiv",
    doi = "10.1016/j.physletb.2006.08.064",
    journal = "Phys. Lett. B",
    volume = "641",
    pages = "278--285",
    year = "2006"
}

@article{Guo:2017jvc,
    author = "Guo, Feng-Kun and Hanhart, Christoph and Mei\ss{}ner, Ulf-G. and Wang, Qian and Zhao, Qiang and Zou, Bing-Song",
    title = "{Hadronic molecules}",
    eprint = "1705.00141",
    archivePrefix = "arXiv",
    primaryClass = "hep-ph",
    doi = "10.1103/RevModPhys.90.015004",
    journal = "Rev. Mod. Phys.",
    volume = "90",
    number = "1",
    pages = "015004",
    year = "2018",
    note = "[Erratum: Rev.Mod.Phys. 94, 029901 (2022)]"
}

@article{Liu:2012zya,
    author = "Liu, Liuming and Orginos, Kostas and Guo, Feng-Kun and Hanhart, Christoph and Meissner, Ulf-G.",
    title = "{Interactions of charmed mesons with light pseudoscalar mesons from lattice QCD and implications on the nature of the $D_{s0}^*(2317)$}",
    eprint = "1208.4535",
    archivePrefix = "arXiv",
    primaryClass = "hep-lat",
    reportNumber = "JLAB-THY-12-1599",
    doi = "10.1103/PhysRevD.87.014508",
    journal = "Phys. Rev. D",
    volume = "87",
    number = "1",
    pages = "014508",
    year = "2013"
}

@article{Lakhina:2006fy,
    author = "Lakhina, Olga and Swanson, Eric S.",
    title = "{A Canonical Ds(2317)?}",
    eprint = "hep-ph/0608011",
    archivePrefix = "arXiv",
    doi = "10.1016/j.physletb.2007.01.075",
    journal = "Phys. Lett. B",
    volume = "650",
    pages = "159--165",
    year = "2007"
}

@article{Guo:2015dha,
    author = "Guo, Zhi-Hui and Mei\ss{}ner, Ulf-G. and Yao, De-Liang",
    title = "{New insights into the $D^{*}_{s0}(2317)$ and other charm scalar mesons}",
    eprint = "1507.03123",
    archivePrefix = "arXiv",
    primaryClass = "hep-ph",
    doi = "10.1103/PhysRevD.92.094008",
    journal = "Phys. Rev. D",
    volume = "92",
    number = "9",
    pages = "094008",
    year = "2015"
}

@article{Guo:2009ct,
    author = "Guo, Feng-Kun and Hanhart, Christoph and Meissner, Ulf-G.",
    title = "{Interactions between heavy mesons and Goldstone bosons from chiral dynamics}",
    eprint = "0901.1597",
    archivePrefix = "arXiv",
    primaryClass = "hep-ph",
    reportNumber = "FZJ-IKP-TH-2009-01, HISKP-TH-09-01",
    doi = "10.1140/epja/i2009-10762-1",
    journal = "Eur. Phys. J. A",
    volume = "40",
    pages = "171--179",
    year = "2009"
}

@article{Guo:2008gp,
    author = "Guo, Feng-Kun and Hanhart, Christoph and Krewald, Siegfried and Meissner, Ulf-G.",
    title = "{Subleading contributions to the width of the D*(s0)(2317)}",
    eprint = "0806.3374",
    archivePrefix = "arXiv",
    primaryClass = "hep-ph",
    reportNumber = "FZJ-IKP(TH)-2008-09, HISKP-TH-08-09",
    doi = "10.1016/j.physletb.2008.07.060",
    journal = "Phys. Lett. B",
    volume = "666",
    pages = "251--255",
    year = "2008"
}

@article{Gamermann:2006nm,
    author = "Gamermann, D. and Oset, E. and Strottman, D. and Vicente Vacas, M. J.",
    title = "{Dynamically generated open and hidden charm meson systems}",
    eprint = "hep-ph/0612179",
    archivePrefix = "arXiv",
    doi = "10.1103/PhysRevD.76.074016",
    journal = "Phys. Rev. D",
    volume = "76",
    pages = "074016",
    year = "2007"
}

@article{Faessler:2007gv,
    author = "Faessler, Amand and Gutsche, Thomas and Lyubovitskij, Valery E. and Ma, Yong-Liang",
    title = "{Strong and radiative decays of the D(s0)*(2317) meson in the DK-molecule picture}",
    eprint = "0705.0254",
    archivePrefix = "arXiv",
    primaryClass = "hep-ph",
    doi = "10.1103/PhysRevD.76.014005",
    journal = "Phys. Rev. D",
    volume = "76",
    pages = "014005",
    year = "2007"
}

@article{Flynn:2007ki,
    author = "Flynn, Jonathan M and Nieves, Juan",
    title = "{Elastic s-wave B pi, D pi, D K and K pi scattering from lattice calculations of scalar form-factors in semileptonic decays}",
    eprint = "hep-ph/0703047",
    archivePrefix = "arXiv",
    doi = "10.1103/PhysRevD.75.074024",
    journal = "Phys. Rev. D",
    volume = "75",
    pages = "074024",
    year = "2007"
}

@article{Bali:2003jv,
    author = "Bali, Gunnar S.",
    title = "{The D+(sJ)(2317): What can the lattice say?}",
    eprint = "hep-ph/0305209",
    archivePrefix = "arXiv",
    doi = "10.1103/PhysRevD.68.071501",
    journal = "Phys. Rev. D",
    volume = "68",
    pages = "071501",
    year = "2003"
}

@article{Dougall:2003hv,
    author = "Dougall, A. and Kenway, R. D. and Maynard, C. M. and McNeile, C.",
    collaboration = "UKQCD",
    title = "{The Spectrum of D(s) mesons from lattice QCD}",
    eprint = "hep-lat/0307001",
    archivePrefix = "arXiv",
    reportNumber = "TRINLAT-03-02, EDINBURGH-2003-10",
    doi = "10.1016/j.physletb.2003.07.017",
    journal = "Phys. Lett. B",
    volume = "569",
    pages = "41--44",
    year = "2003"
}

@article{Mohler:2012na,
    author = "Mohler, Daniel and Prelovsek, Sasa and Woloshyn, R. M.",
    title = "{$D \pi$ scattering and $D$ meson resonances from lattice QCD}",
    eprint = "1208.4059",
    archivePrefix = "arXiv",
    primaryClass = "hep-lat",
    doi = "10.1103/PhysRevD.87.034501",
    journal = "Phys. Rev. D",
    volume = "87",
    number = "3",
    pages = "034501",
    year = "2013"
}

@article{Mohler:2013rwa,
    author = "Mohler, Daniel and Lang, C. B. and Leskovec, Luka and Prelovsek, Sasa and Woloshyn, R. M.",
    title = "{$D_{s0}^*(2317)$ Meson and $D$-Meson-Kaon Scattering from Lattice QCD}",
    eprint = "1308.3175",
    archivePrefix = "arXiv",
    primaryClass = "hep-lat",
    reportNumber = "FERMILAB-PUB-13-318-T",
    doi = "10.1103/PhysRevLett.111.222001",
    journal = "Phys. Rev. Lett.",
    volume = "111",
    number = "22",
    pages = "222001",
    year = "2013"
}

@article{Cheung:2020mql,
    author = "Cheung, Gavin K. C. and Thomas, Christopher E. and Wilson, David J. and Moir, Graham and Peardon, Michael and Ryan, Sin\'ead M.",
    collaboration = "Hadron Spectrum",
    title = "{DK I = 0,$ D\overline{K} $I = 0, 1 scattering and the $ {D}_{s0}^{\ast } $(2317) from lattice QCD}",
    eprint = "2008.06432",
    archivePrefix = "arXiv",
    primaryClass = "hep-lat",
    doi = "10.1007/JHEP02(2021)100",
    journal = "JHEP",
    volume = "02",
    pages = "100",
    year = "2021"
}

@article{Liu:2023uly,
    author = "Liu, Zhi-Wei and Lu, Jun-Xu and Geng, Li-Sheng",
    title = "{Study of the DK interaction with femtoscopic correlation functions}",
    eprint = "2302.01046",
    archivePrefix = "arXiv",
    primaryClass = "hep-ph",
    doi = "10.1103/PhysRevD.107.074019",
    journal = "Phys. Rev. D",
    volume = "107",
    number = "7",
    pages = "074019",
    year = "2023"
}

@article{Albaladejo:2023pzq,
    author = "Albaladejo, Miguel and Nieves, Juan and Ruiz-Arriola, Enrique",
    title = "{Femtoscopic signatures of the lightest S-wave scalar open-charm mesons}",
    eprint = "2304.03107",
    archivePrefix = "arXiv",
    primaryClass = "hep-ph",
    doi = "10.1103/PhysRevD.108.014020",
    journal = "Phys. Rev. D",
    volume = "108",
    number = "1",
    pages = "014020",
    year = "2023"
}

@article{Ikeno:2023ojl,
    author = "Ikeno, Natsumi and Toledo, Genaro and Oset, Eulogio",
    title = "{Model independent analysis of femtoscopic correlation functions: An application to the Ds0\textasteriskcentered{}(2317)}",
    eprint = "2305.16431",
    archivePrefix = "arXiv",
    primaryClass = "hep-ph",
    doi = "10.1016/j.physletb.2023.138281",
    journal = "Phys. Lett. B",
    volume = "847",
    pages = "138281",
    year = "2023"
}

@article{Torres-Rincon:2023qll,
    author = "Torres-Rincon, Juan M. and Ramos, \`Angels and Tolos, Laura",
    title = "{Femtoscopy of D mesons and light mesons upon unitarized effective field theories}",
    eprint = "2307.02102",
    archivePrefix = "arXiv",
    primaryClass = "hep-ph",
    doi = "10.1103/PhysRevD.108.096008",
    journal = "Phys. Rev. D",
    volume = "108",
    number = "9",
    pages = "096008",
    year = "2023"
}

@article{Albaladejo:2016lbb,
    author = "Albaladejo, Miguel and Fernandez-Soler, Pedro and Guo, Feng-Kun and Nieves, Juan",
    title = "{Two-pole structure of the $D^\ast_0(2400)$}",
    eprint = "1610.06727",
    archivePrefix = "arXiv",
    primaryClass = "hep-ph",
    doi = "10.1016/j.physletb.2017.02.036",
    journal = "Phys. Lett. B",
    volume = "767",
    pages = "465--469",
    year = "2017"
}

@article{Albaladejo:2018mhb,
    author = "Albaladejo, Miguel and Fernandez-Soler, Pedro and Nieves, Juan and Ortega, Pablo G.",
    title = "{Contribution of constituent quark model $c\bar{s}$ states to the dynamics of the $D_{s0}^*(2317)$ and $D_{s1}(2460)$ resonances}",
    eprint = "1805.07104",
    archivePrefix = "arXiv",
    primaryClass = "hep-ph",
    doi = "10.1140/epjc/s10052-018-6176-3",
    journal = "Eur. Phys. J. C",
    volume = "78",
    number = "9",
    pages = "722",
    year = "2018"
}

@article{Ortega:2016mms,
    author = "Ortega, Pablo G. and Segovia, Jorge and Entem, David R. and Fernandez, Francisco",
    title = "{Molecular components in P-wave charmed-strange mesons}",
    eprint = "1603.07000",
    archivePrefix = "arXiv",
    primaryClass = "hep-ph",
    doi = "10.1103/PhysRevD.94.074037",
    journal = "Phys. Rev. D",
    volume = "94",
    number = "7",
    pages = "074037",
    year = "2016"
}

@article{Albaladejo:2016hae,
    author = "Albaladejo, Miguel and Jido, Daisuke and Nieves, Juan and Oset, Eulogio",
    title = "{$D^*_{s0}(2317)$ and $\textit{DK}$ scattering in B decays from BaBar and LHCb data}",
    eprint = "1604.01193",
    archivePrefix = "arXiv",
    primaryClass = "hep-ph",
    doi = "10.1140/epjc/s10052-016-4144-3",
    journal = "Eur. Phys. J. C",
    volume = "76",
    number = "6",
    pages = "300",
    year = "2016"
}

@article{Lang:2014yfa,
    author = "Lang, C. B. and Leskovec, Luka and Mohler, Daniel and Prelovsek, Sasa and Woloshyn, R. M.",
    title = "{Ds mesons with DK and D*K scattering near threshold}",
    eprint = "1403.8103",
    archivePrefix = "arXiv",
    primaryClass = "hep-lat",
    reportNumber = "FERMILAB-PUB-14-063-T",
    doi = "10.1103/PhysRevD.90.034510",
    journal = "Phys. Rev. D",
    volume = "90",
    number = "3",
    pages = "034510",
    year = "2014"
}

@article{Bali:2017pdv,
    author = {Bali, Gunnar S. and Collins, Sara and Cox, Antonio and Sch\"afer, Andreas},
    title = "{Masses and decay constants of the $D_{s0}^*(2317)$ and $D_{s1}(2460)$ from $N_f=2$ lattice QCD close to the physical point}",
    eprint = "1706.01247",
    archivePrefix = "arXiv",
    primaryClass = "hep-lat",
    doi = "10.1103/PhysRevD.96.074501",
    journal = "Phys. Rev. D",
    volume = "96",
    number = "7",
    pages = "074501",
    year = "2017"
}

@article{Molina:2009zeg,
    author = "Molina, R. and Gamermann, D. and Oset, E. and Tolos, L.",
    title = "{Charm and hidden charm scalar mesons in the nuclear medium}",
    eprint = "0806.3711",
    archivePrefix = "arXiv",
    primaryClass = "nucl-th",
    doi = "10.1140/epja/i2009-10853-y",
    journal = "Eur. Phys. J. A",
    volume = "42",
    pages = "31--42",
    year = "2009"
}

@article{Montana:2020lfi,
    author = "Monta\~na, Gl\`oria and Ramos, \`Angels and Tolos, Laura and Torres-Rincon, Juan M.",
    title = "{Impact of a thermal medium on $D$ mesons and their chiral partners}",
    eprint = "2001.11877",
    archivePrefix = "arXiv",
    primaryClass = "hep-ph",
    doi = "10.1016/j.physletb.2020.135464",
    journal = "Phys. Lett. B",
    volume = "806",
    pages = "135464",
    year = "2020"
}

@article{Montana:2020vjg,
    author = "Monta{\~n}a, Gl{\`o}ria and Ramos, {\`A}ngels and Tolos, Laura and Torres-Rincon, Juan M.",
    title = "{Pseudoscalar and vector open-charm mesons at finite temperature}",
    eprint = "2007.12601",
    archivePrefix = "arXiv",
    primaryClass = "hep-ph",
    doi = "10.1103/PhysRevD.102.096020",
    journal = "Phys. Rev. D",
    volume = "102",
    number = "9",
    pages = "096020",
    year = "2020"
}

@article{Montana:2023sft,
    author = "Montana, Gloria and Ramos, Angels and Tolos, Laura and Torres-Rincon, Juan M.",
    title = "{Recent progress on in-medium properties of heavy mesons from finite-temperature EFTs}",
    eprint = "2307.03640",
    archivePrefix = "arXiv",
    primaryClass = "hep-ph",
    reportNumber = "JLAB-THY-23-3874",
    doi = "10.3389/fphy.2023.1250939",
    journal = "Front. in Phys.",
    volume = "11",
    pages = "1250939",
    year = "2023"
}

@article{Montesinos:2024uhq,
    author = "Montesinos, Victor and Albaladejo, Miguel and Nieves, Juan and Tolos, Laura",
    title = "{Charge-conjugation asymmetry and molecular content: The Ds0{\textasteriskcentered}(2317){\ensuremath{\pm}} in matter}",
    eprint = "2403.00451",
    archivePrefix = "arXiv",
    primaryClass = "hep-ph",
    doi = "10.1016/j.physletb.2024.138656",
    journal = "Phys. Lett. B",
    volume = "853",
    pages = "138656",
    year = "2024"
}

@article{Montesinos:2023qbx,
    author = "Montesinos, Victor and Albaladejo, Miguel and Nieves, Juan and Tolos, Laura",
    title = "{Properties of the Tcc(3875)+ and Tc{\textasciimacron}c{\textasciimacron}(3875){\ensuremath{-}} and their heavy-quark spin partners in nuclear matter}",
    eprint = "2306.17673",
    archivePrefix = "arXiv",
    primaryClass = "hep-ph",
    doi = "10.1103/PhysRevC.108.035205",
    journal = "Phys. Rev. C",
    volume = "108",
    number = "3",
    pages = "035205",
    year = "2023"
}

@article{Albaladejo:2021cxj,
    author = "Albaladejo, M. and Nieves, J. M. and Tolos, L.",
    title = "{DD{\textasciimacron}* scattering and {\ensuremath{\chi}}c1(3872) in nuclear matter}",
    eprint = "2102.08589",
    archivePrefix = "arXiv",
    primaryClass = "hep-ph",
    reportNumber = "JLAB-THY-21-3321",
    doi = "10.1103/PhysRevC.104.035203",
    journal = "Phys. Rev. C",
    volume = "104",
    number = "3",
    pages = "035203",
    year = "2021"
}

@article{Tolos:2007vh,
    author = "Tolos, Laura and Ramos, Angels and Mizutani, Tetsuro",
    title = "{Open charm in nuclear matter at finite temperature}",
    eprint = "0710.2684",
    archivePrefix = "arXiv",
    primaryClass = "nucl-th",
    doi = "10.1103/PhysRevC.77.015207",
    journal = "Phys. Rev. C",
    volume = "77",
    pages = "015207",
    year = "2008"
}

@article{Gasser:1986vb,
    author = "Gasser, J. and Leutwyler, H.",
    title = "{Light Quarks at Low Temperatures}",
    reportNumber = "BUTP-86/19-BERN",
    doi = "10.1016/0370-2693(87)90492-8",
    journal = "Phys. Lett. B",
    volume = "184",
    pages = "83--88",
    year = "1987"
}

@article{Gerber:1988tt,
    author = "Gerber, P. and Leutwyler, H.",
    title = "{Hadrons Below the Chiral Phase Transition}",
    reportNumber = "BUTP-88/30-BERN",
    doi = "10.1016/0550-3213(89)90349-0",
    journal = "Nucl. Phys. B",
    volume = "321",
    pages = "387--429",
    year = "1989"
}

@article{Kodama:1995kj,
    author = "Kodama, Nobuaki and Oka, Makoto",
    title = "{The Pion at finite temperature}",
    eprint = "hep-ph/9510401",
    archivePrefix = "arXiv",
    reportNumber = "TIT-HEP-306-NP",
    doi = "10.1016/0375-9474(95)00497-1",
    journal = "Nucl. Phys. A",
    volume = "601",
    pages = "304--318",
    year = "1996"
}

@article{Bochkarev:1995gi,
    author = "Bochkarev, Alexander and Kapusta, Joseph I.",
    title = "{Chiral symmetry at finite temperature: Linear versus nonlinear sigma models}",
    eprint = "hep-ph/9602405",
    archivePrefix = "arXiv",
    reportNumber = "NUC-MINN-95-25-T, TPI-MINN-95-11-T, HEP-MINN-TH-1342",
    doi = "10.1103/PhysRevD.54.4066",
    journal = "Phys. Rev. D",
    volume = "54",
    pages = "4066--4079",
    year = "1996"
}

@article{Kim:2003tp,
    author = "Kim, Hung-chong and Oka, Makoto",
    title = "{Update on pion weak decay constants in nuclear matter}",
    eprint = "hep-ph/0301227",
    archivePrefix = "arXiv",
    doi = "10.1016/S0375-9474(03)01004-2",
    journal = "Nucl. Phys. A",
    volume = "720",
    pages = "368--381",
    year = "2003"
}

@article{Kaiser:2007nv,
    author = "Kaiser, N. and de Homont, P. and Weise, W.",
    title = "{In-medium chiral condensate beyond linear density approximation}",
    eprint = "0711.3154",
    archivePrefix = "arXiv",
    primaryClass = "nucl-th",
    doi = "10.1103/PhysRevC.77.025204",
    journal = "Phys. Rev. C",
    volume = "77",
    pages = "025204",
    year = "2008"
}

@article{Jido:2008bk,
    author = "Jido, D. and Hatsuda, T. and Kunihiro, T.",
    title = "{In-medium Pion and Partial Restoration of Chiral Symmetry}",
    eprint = "0805.4453",
    archivePrefix = "arXiv",
    primaryClass = "nucl-th",
    reportNumber = "YITP-07-23",
    doi = "10.1016/j.physletb.2008.10.034",
    journal = "Phys. Lett. B",
    volume = "670",
    pages = "109--113",
    year = "2008"
}

@article{Alarcon:2011zs,
    author = "Alarcon, J. M. and Martin Camalich, J. and Oller, J. A.",
    title = "{The chiral representation of the $\pi N$ scattering amplitude and the pion-nucleon sigma term}",
    eprint = "1110.3797",
    archivePrefix = "arXiv",
    primaryClass = "hep-ph",
    doi = "10.1103/PhysRevD.85.051503",
    journal = "Phys. Rev. D",
    volume = "85",
    pages = "051503",
    year = "2012"
}

@article{Goda:2013npa,
    author = "Goda, Soichiro and Jido, Daisuke",
    title = "{Pion properties at finite nuclear density based on in-medium chiral perturbation theory}",
    eprint = "1312.0832",
    archivePrefix = "arXiv",
    primaryClass = "nucl-th",
    doi = "10.1093/ptep/ptu023",
    journal = "PTEP",
    volume = "2014",
    number = "3",
    pages = "033D03",
    year = "2014"
}

@article{Waas:1997pe,
    author = "Waas, T. and Weise, W.",
    title = "{S wave interactions of anti-K and eta mesons in nuclear matter}",
    doi = "10.1016/S0375-9474(97)00487-9",
    journal = "Nucl. Phys. A",
    volume = "625",
    pages = "287--306",
    year = "1997"
}

@article{Inoue:2002xw,
    author = "Inoue, T. and Oset, E.",
    title = "{Eta in the nuclear medium within a chiral unitary approach}",
    eprint = "hep-ph/0205028",
    archivePrefix = "arXiv",
    doi = "10.1016/S0375-9474(02)01167-3",
    journal = "Nucl. Phys. A",
    volume = "710",
    pages = "354--370",
    year = "2002"
}

@article{Post:2003hu,
    author = "Post, M. and Leupold, S. and Mosel, U.",
    title = "{Hadronic spectral functions in nuclear matter}",
    eprint = "nucl-th/0309085",
    archivePrefix = "arXiv",
    doi = "10.1016/j.nuclphysa.2004.05.016",
    journal = "Nucl. Phys. A",
    volume = "741",
    pages = "81--148",
    year = "2004"
}

@article{Jimenez-Tejero:2009cyn,
    author = "Jimenez-Tejero, C. E. and Ramos, A. and Vidana, I.",
    title = "{Dynamically generated open charmed baryons beyond the zero range approximation}",
    eprint = "0907.5316",
    archivePrefix = "arXiv",
    primaryClass = "hep-ph",
    doi = "10.1103/PhysRevC.80.055206",
    journal = "Phys. Rev. C",
    volume = "80",
    pages = "055206",
    year = "2009"
}

@article{Jimenez-Tejero:2011dif,
    author = "Jimenez-Tejero, C. E. and Ramos, A. and Tolos, L. and Vidana, I.",
    title = "{Open charm meson in nuclear matter at finite temperature beyond the zero range approximation}",
    eprint = "1102.4786",
    archivePrefix = "arXiv",
    primaryClass = "hep-ph",
    doi = "10.1103/PhysRevC.84.015208",
    journal = "Phys. Rev. C",
    volume = "84",
    pages = "015208",
    year = "2011"
}

@article{Lutz:2005vx,
    author = "Lutz, M. F. M. and Korpa, C. L.",
    title = "{Open-charm systems in cold nuclear matter}",
    eprint = "nucl-th/0510006",
    archivePrefix = "arXiv",
    doi = "10.1016/j.physletb.2005.11.046",
    journal = "Phys. Lett. B",
    volume = "633",
    pages = "43--48",
    year = "2006"
}

@article{Hofmann:2005sw,
    author = "Hofmann, J. and Lutz, M. F. M.",
    title = "{Coupled-channel study of crypto-exotic baryons with charm}",
    eprint = "hep-ph/0507071",
    archivePrefix = "arXiv",
    doi = "10.1016/j.nuclphysa.2005.08.022",
    journal = "Nucl. Phys. A",
    volume = "763",
    pages = "90--139",
    year = "2005"
}

@article{Garcia-Gonzales:2026dnn,
    author = "Garcia-Gonzales, E. E. and Magas, V. K. and Ramos, A.",
    title = "{Comprehensive study of hidden charm pentaquarks with an improved unitarization method}",
    eprint = "2605.21205",
    archivePrefix = "arXiv",
    primaryClass = "hep-ph",
    month = "5",
    year = "2026"
}

@article{Garcia-Recio:2008rjt,
    author = "Garcia-Recio, C. and Magas, V. K. and Mizutani, T. and Nieves, J. and Ramos, A. and Salcedo, L. L. and Tolos, L.",
    title = "{The s-wave charmed baryon resonances from a coupled-channel approach with heavy quark symmetry}",
    eprint = "0807.2969",
    archivePrefix = "arXiv",
    primaryClass = "hep-ph",
    doi = "10.1103/PhysRevD.79.054004",
    journal = "Phys. Rev. D",
    volume = "79",
    pages = "054004",
    year = "2009"
}

@article{Tolos:2009nn,
    author = "Tolos, L. and Garcia-Recio, C. and Nieves, J.",
    title = "{The Properties of D and D* mesons in the nuclear medium}",
    eprint = "0905.4859",
    archivePrefix = "arXiv",
    primaryClass = "nucl-th",
    doi = "10.1103/PhysRevC.80.065202",
    journal = "Phys. Rev. C",
    volume = "80",
    pages = "065202",
    year = "2009"
}

@article{Geng:2010vw,
    author = "Geng, L. S. and Kaiser, N. and Martin-Camalich, J. and Weise, W.",
    title = "{Low-energy interactions of Nambu-Goldstone bosons with $D$ mesons in covariant chiral perturbation theory}",
    eprint = "1008.0383",
    archivePrefix = "arXiv",
    primaryClass = "hep-ph",
    doi = "10.1103/PhysRevD.82.054022",
    journal = "Phys. Rev. D",
    volume = "82",
    pages = "054022",
    year = "2010"
}

@article{Abreu:2011ic,
    author = "Abreu, Luciano M. and Cabrera, Daniel and Llanes-Estrada, Felipe J. and Torres-Rincon, Juan M.",
    title = "{Charm diffusion in a pion gas implementing unitarity, chiral and heavy quark symmetries}",
    eprint = "1104.3815",
    archivePrefix = "arXiv",
    primaryClass = "hep-ph",
    doi = "10.1016/j.aop.2011.06.006",
    journal = "Annals Phys.",
    volume = "326",
    pages = "2737--2772",
    year = "2011"
}

@article{Lutz:2007sk,
    author = "Lutz, Matthias F. M. and Soyeur, Madeleine",
    title = "{Radiative and isospin-violating decays of D(s)-mesons in the hadrogenesis conjecture}",
    eprint = "0710.1545",
    archivePrefix = "arXiv",
    primaryClass = "hep-ph",
    reportNumber = "DAPNIA-07-147",
    doi = "10.1016/j.nuclphysa.2008.09.003",
    journal = "Nucl. Phys. A",
    volume = "813",
    pages = "14--95",
    year = "2008"
}

@article{Guo:2018tjx,
    author = "Guo, Zhi-Hui and Liu, Liuming and Mei{\ss}ner, Ulf-G and Oller, J. A. and Rusetsky, A.",
    title = "{Towards a precise determination of the scattering amplitudes of the charmed and light-flavor pseudoscalar mesons}",
    eprint = "1811.05585",
    archivePrefix = "arXiv",
    primaryClass = "hep-ph",
    doi = "10.1140/epjc/s10052-018-6518-1",
    journal = "Eur. Phys. J. C",
    volume = "79",
    number = "1",
    pages = "13",
    year = "2019"
}

@article{Matsubara:1955ws,
    author = "Matsubara, Takeo",
    title = "{A New approach to quantum statistical mechanics}",
    doi = "10.1143/PTP.14.351",
    journal = "Prog. Theor. Phys.",
    volume = "14",
    pages = "351--378",
}

@book{Kapusta:2006pm,
    author = "Kapusta, J. I. and Gale, Charles",
    title = "{Finite-temperature field theory: Principles and applications}",
    doi = "10.1017/CBO9780511535130",
    isbn = "978-0-521-17322-3, 978-0-521-82082-0, 978-0-511-22280-1",
    publisher = "Cambridge University Press",
    series = "Cambridge Monographs on Mathematical Physics",
    year = "2011"
}

@article{Tolos:2008di,
    author = "Tolos, L. and Cabrera, D. and Ramos, A.",
    title = "{Strange mesons in nuclear matter at finite temperature}",
    eprint = "0807.2947",
    archivePrefix = "arXiv",
    primaryClass = "nucl-th",
    doi = "10.1103/PhysRevC.78.045205",
    journal = "Phys. Rev. C",
    volume = "78",
    pages = "045205",
    year = "2008"
}

@article{Tolos:2020aln,
    author = "Tolos, Laura and Fabbietti, Laura",
    title = "{Strangeness in Nuclei and Neutron Stars}",
    eprint = "2002.09223",
    archivePrefix = "arXiv",
    primaryClass = "nucl-ex",
    doi = "10.1016/j.ppnp.2020.103770",
    journal = "Prog. Part. Nucl. Phys.",
    volume = "112",
    pages = "103770",
    year = "2020"
}

@article{Oller:1998hw,
    author = "Oller, J. A. and Oset, E. and Pelaez, J. R.",
    title = "{Meson meson interaction in a nonperturbative chiral approach}",
    eprint = "hep-ph/9804209",
    archivePrefix = "arXiv",
    reportNumber = "SLAC-PUB-7787",
    doi = "10.1103/PhysRevD.59.074001",
    journal = "Phys. Rev. D",
    volume = "59",
    pages = "074001",
    year = "1999",
    note = "[Erratum: Phys.Rev.D 60, 099906 (1999), Erratum: Phys.Rev.D 75, 099903 (2007)]"
}

\end{document}